\documentclass{sig-alternate}
\usepackage{cite}
\usepackage{amsmath,amssymb,amsfonts}
\usepackage{mathptmx}
\usepackage{comment}
\usepackage{bm}
\usepackage{algorithmic}
\usepackage{graphicx}
\usepackage{textcomp}
\usepackage{fancyhdr}
\usepackage[hyphens]{url}
\usepackage[normalem]{ulem}
\usepackage[bookmarks=true,breaklinks=true,colorlinks,linkcolor=black,citecolor=blue,urlcolor=black]{hyperref}
\hypersetup{
    unicode=false,          
    pdftoolbar=true,        
    pdfmenubar=true,        
    pdffitwindow=false,     
    pdftitle={BARISTA},    
    pdfauthor={Mithuna Thottethodi},     
    pdfsubject={Barrier-Free large-Scale Sparse Tensor Accelerator},   
    pdfcreator={Ashish Gondimalla},   
    pdfproducer={},         
    pdfkeywords={Accelerators,} {Convolutional neural networks,} {Sparse Tensors}, 
    pdfnewwindow=true,      
    colorlinks=false,       
    linkcolor=red,          
    citecolor=green,        
    filecolor=magenta,      
    urlcolor=cyan           
}

\def\BibTeX{{\rm B\kern-.05em{\sc i\kern-.025em b}\kern-.08em
    T\kern-.1667em\lower.7ex\hbox{E}\kern-.125emX}}

\title{Barrier-Free Large-Scale Sparse Tensor Accelerator (BARISTA) For Convolutional Neural Networks}
\author{Ashish Gondimalla, Sree Charan Gundabolu, T. N. Vijaykumar, and Mithuna Thottethodi\\
School of Electrical and Computer Engineering, Purdue University  \\
Email: {\tt agondima@purdue.edu,sgundabo@purdue.edu,vijay@ecn.purdue.edu,mithuna@purdue.edu}
}

\usepackage{graphicx}
\usepackage{ifthen}
\usepackage{calc}
\usepackage{pifont}
\usepackage{color}
\usepackage{soul}
\usepackage{fancyhdr}
\usepackage{xspace}
\usepackage{multirow}
\usepackage[table]{xcolor} 
\definecolor{lightgray}{gray}{0.9}
\newenvironment{stripetabular}{\rowcolors{2}{white}{lightgray}\tabular}{\endtabular}

\newcounter{hours}
\newcounter{minutes}

\newcommand{\name}{{\em BARISTA}\xspace}

\newboolean{timeofmake}
\setboolean{timeofmake}{false}

\newcommand{\ignore}[1]{}

\newcommand{\dontinclude}[1]{ }

\newcommand{\putsec}[2]{\vspace{-0.05in}\section{#2}\label{sec:#1}\vspace{-0.01in}}
\newcommand{\putsubsec}[2]{\vspace{-0.05in}\subsection{#2}\label{sec:#1}\vspace{-0.01in}}
\newcommand{\putsubsubsec}[2]{\vspace{0.05in}\subsubsection{#2}\label{sec:#1}\vspace{0.01in}}

\newcommand{\tabput}[3]{
\begin{table}[t]
\footnotesize 
\vspace{-0.1in}
\caption{\footnotesize\sf #3 \label{tab:#1}}
\begin{center}
{
#2
}
\end{center}
\vspace{-0.25in}
\end{table}
}

\newcommand{\figput}[4][1.0\linewidth]{
\begin{figure}[t]
\begin{minipage}{\linewidth}
\footnotesize 
\begin{center}
\includegraphics[clip, width=#1]{figures/#2}
\end{center}
\vspace{-0.25in}
\caption{#4 \label{fig:#2}}
\vspace{-0.2in}
\end{minipage}
\end{figure}
}

\newcommand{\figputW}[4][1.0\linewidth]{
\begin{figure*}[t]
\begin{minipage}{\linewidth}
\footnotesize 
\begin{center}
\includegraphics[clip, width=#1]{figures/#2}
\end{center}
\vspace{-0.25in}
\caption{#4 \label{fig:#2}}
\vspace{-0.2in}
\end{minipage}
\end{figure*}
}

\newcommand{\figref}[1]{Figure~\ref{fig:#1}}
\newcommand{\tabref}[1]{Table~\ref{tab:#1}}
\newcommand{\secref}[1]{Section~\ref{sec:#1}}

\begin{document}
\maketitle
\thispagestyle{plain}
\pagestyle{plain}

\begin{abstract}   

Convolutional neural networks (CNNs) are emerging as powerful tools for visual recognition.  Recent architecture proposals for sparse CNNs exploit natural and transformed zeros in the feature maps and filters for performance and energy without losing accuracy. Sparse architectures that exploit two-sided sparsity in both feature maps and filters  have been studied only at small scales (e.g., 1K multiply-accumulate units (MACs)). However, to realize their advantages in full,
the sparse architectures have to be scaled up to levels of the dense architectures (e.g., 32K MACs in the TPU). Such scaling is challenging because achieving reuse through broadcasts incurs implicit barrier cost which raises the inter-related issues of load imbalance, 
buffering, and on-chip bandwidth demand.  
SparTen, a previous scheme,  addresses one aspect of load balancing but not other aspects,  nor the other issues of buffering and bandwidth.
To that end, we propose the {\em barrier-free large-scale sparse tensor accelerator (BARISTA)}. 
\name (1) is the first architecture for scaling up sparse CNN accelerators; 
(2) reduces on-chip bandwidth demand by  telescoping request-combining  the input map requests and  snarfing  the filter requests; (3) reduces buffering via basic buffer sharing  and avoids the ensuing barriers between consecutive  input maps by coloring the output buffers;  (4) load balances  intra-filter work  via  dynamic round-robin work assignment; and (5) employs hierarchical buffering which achieves high cache bandwidth via a few, wide, shared buffers and low buffering via narrower, private buffers at the compute.
Our simulations show that, on average, \name  performs  5.4x, 2.2x, 1.7x, 2.5x better 
than a dense, a one-sided, a naively-scaled two-sided, and an iso-area two-sided architecture, respectively.  {Using 45-nm technology,}
ASIC synthesis of our RTL implementation for four clusters of 8K MACs each  reports 1 GHz clock speed, 213 mm$^2$ area and 170 W power.

\end{abstract}

\putsec{intro}{Introduction}

Convolutional neural networks (CNNs) are resulting in highly-accurate visual recognition ~\cite{lecun,alexnet,vggnet,resnet}.  While CNNs involve heavy computation and large intermediate data, there is significant sparsity in CNN inference 
caused by  the Rectifier Linear Unit (ReLU) which converts negative values in the feature maps to zeros.  Further,  previous work actively increases sparsity by pruning filters followed by retraining to maintain accuracy~\cite{Dally-NIPS,Dally-ICLR}. The combined feature map and filter sparsity leads to 2-3x memory size reduction and 4-9x compute reduction. Being hard-wired for regularity in compute and data access, {\em dense architectures} like the GPU and TPU cannot efficiently exploit sparsity.  
Past architecture proposals for CNN inference exploit {\em one-sided} sparsity in either the feature maps~\cite{cnvlutin,cambriconx,EIE}  or the filters~\cite{cambricons}, or {\em two-sided} sparsity~\cite{scnn, sparten}.

Because one-sided architectures  do not exploit the full opportunity afforded by sparsity, 
we focus on two-sided sparsity.  
The two-sided sparse accelerators' performance and energy advantages over the dense counterparts (e.g.,  GPUs or the TPU~\cite{tpu}) have been shown {\em only} at small scales (e.g., 1K multiply-add units (MACs) in SparTen and SCNN). While such small scales may be relevant for some edge  contexts, other contexts 
require full, large-scale acceleration (e.g., 32K MACs in the TPU). For example, Tesla's Full Self-Driving delivers 144 trillion operations per second (TOPS)~\cite{TeslaFSD}.
Similarly, Nvidia's Drive platforms's compute continues to scale up from 320 TOPS (Pegasus, 2017) to 400 TOPS (Orin, 2019)~\cite{agxorin}. Of this 144 (or 320) TOPS, we envision a
 portion (say 32 TOPS)  for CNN inference (the remaining TOPS may be for non-CNN tasks).
At any scale,  sparsity's  advantages are realized fully only if the sparse accelerators are scaled up as much as the dense accelerators (e.g., at large scale, the sparse accelerators must also keep  32K MACs busy). 
However, 
scaling up the sparse accelerators  faces  the fundamentally opposing forces of load imbalance and reuse, making high performance and low energy  challenging (e.g., load-balancing all the 32K MACs). 
Our notion of scaling up is purely in terms of more MACs. The `scale-up' (larger MAC clusters) versus `scale-out' (more small clusters) considerations in systolic architectures~\cite{scaleout}are not relevant to our architecture, as we explain later.

To achieve good reuse while controlling buffering and on-chip bandwidth demand, current sparse accelerators  employ broadcasts which impose (implicit) barriers. For instance,  each filter is broadcast just once to be reused by all the  input feature maps imposing an implicit barrier at each broadcast to avoid excessive buffering  (the broadcast may use  a physical bus or pipelined links). 
The barriers incur load imbalance cost which is modest at small scales but  significant at large scales {(e.g., our results show that eliminating the barrier cost improves performance by 72\% for 32K MACs).} While the dense accelerators are load balanced naturally, the sparse accelerators' load imbalance  significantly diminishes the sparsity advantages at such extreme scales.  Avoiding 
the barrier cost by having each input feature map refetch the same filter would result in inordinately high  on-chip bandwidth demand (e.g., each filter would be  refetched 64 times from the cache). Alternatively, separately buffering the filter for each asynchronous  input feature map 
would result in exploding buffering requirements (for a set of filters, a sparser input map would systematically diverge from  a denser input map, requiring ever-growing buffering). Even without such separate buffering, SparTen assumes 1 KB buffering per MAC which would amount to 32 MB of buffering for 32K MACs.
Thus,  avoiding the barrier overhead at large scale exacerbates the  {\em inter-related} issues of  buffering and bandwidth demand. SparTen handles one aspect of load balancing  but not other aspects as we explain below, nor the other issues of buffering and bandwidth.  

We fundamentally change the two-sided sparse architecture to be {\em barrier-free at scale} via 
our proposal, called the {\em barrier free large-scale sparse tensor accelerator} (\name). 



Previous architectures' {\em clusters}  are small (e.g., 32 MACs  per cluster) limiting the within-cluster reuse; an input map is fetched once and broadcast within a  cluster but the filters are refetched across the clusters 
(inter-cluster broadcast would expose load imbalance).
To achieve good within-cluster reuse, \name employs  large clusters each of which is organized as a three-dimensional grid of {\em filter group rows (FGRs)} and  {\em input feature map group columns  (IFGCs)}  intersecting at 
{\em nodes} which comprise {\em multiple processing elements (PEs)} (e.g., 64 FGRs, 32 IFGCs, and  4 PEs per node  for a total of $64*32*4$ = 8K MACs per cluster).
Each FGR and each IFGC reuses a filter with  multiple input feature maps and an input map with multiple filters, respectively. 
To reduce buffering \name employs basic buffer sharing among the PEs of a node. 
Each FGR and IFGC works independently of the others (i.e., no broadcasts), not in a systolic fashion, and can emulate scaled-out small clusters,  obviating  scale-out considerations. In general, what matters is the granularity of the broadcasts; not the cluster size. Broadcasts to many PEs (within one large cluster or across many small clusters) expose  load imbalance or incur  buffering, whereas broadcasts to fewer PEs (in a subset of a large cluster or a small cluster) or no broadcasts at all incur on-chip bandwidth.

\name makes the following contributions which have a common theme that due to the extreme scale, they are in software or use simple hardware,
and yet bring \name to within 6\% of ideal performance: 

\begin{itemize}

\item Unlike previous schemes, \name is a barrier-free architecture. While there is a plethora of work on dense and sparse CNN accelerators, \name is the first work to scale up sparse CNN accelerators and address the inter-related issues of on-chip bandwidth, load imbalance and buffering.

\item {\bf On-chip bandwidth:} To reduce the  bandwidth demand  without  broadcasts, we make the key observation that if the filters and input feature maps are reasonably load-balanced using the schemes below, 
then most  of the nodes in an FGR would be working on the same input maps.
This in-sync progress  means that an FGR's nodes would request the same input maps at about the same time {\em even without (implicit) barriers}. 
A majority of the fastest nodes  stray  only gradually from each other followed in time by a smaller
group of slower nodes  which are in turn followed by an even smaller group of even slower nodes, and so on.
Accordingly, \name  {\em request combines} {\em telescoping} numbers of input map requests, instead of the same numbers of requests, to prevent performance loss due to the straying  (e.g., out of 64 requests, \name combines the first 48 requests, the next 12, the next  2, leaving the last two uncombined). With reasonable buffering, \name cuts the refetch count from 58  to 7. Filters, being static unlike input maps, are offline load-balanced using a variant of Greedy Balancing~\cite{sparten} where the filters are ordered by density so that filters processed together are of similar density.
Therefore, filters employ simple snarfing, where one request opportunistically places the response in other requesters' buffers subject to buffer availability, instead of telescoping request combining.

\item {\bf Avoiding barriers between input feature maps:} 
While filters are read-only and amenable to pre-processing for load balancing, feature maps are not.  Multiple PEs per node (a) share the node's buffers which reduces buffering and (b) enable fewer IFGCs with lower work variation which improves load imbalance. 
While each node processes multiple input maps per filter for filter reuse,  a barrier among a node's PEs between consecutive input  maps would expose load imbalance across the PEs. To avoid this barrier,
\name \emph{colors} the output buffers so that  a PE  can start processing the next input map even though another PE has not completed the  previous map. The coloring separates the maps' {\em concurrent} partial products into different output buffers.

\item {\bf Intra-filter load balance:} The PEs statically partition the filter assigned to the parent node avoiding complex work-assignment hardware such as a dynamic task queue. While the PEs help  buffering and  IFGC load balance,  the  static partitioning may induce systematic load imbalance across the PEs. For instance, a PE that is assigned a  dense part of a filter for several input maps over time  would lag another PE that is assigned a sparser part. Instead, \name employs {\em dynamic round-robin assignment of filter parts} to load-balance the PEs across input maps. 

\item {\bf Hierarchical buffering:} Wide buffers enable high cache bandwidth via wide fetches but increase the amount of buffering. Instead, \name employs hierarchical buffering which achieves high cache bandwidth by fetching into shared buffers and low buffering via narrower, private buffers at the compute.

\end{itemize}

{While SparTen alleviates }{\em {inter}}{-filter load imbalance via Greedy Balancing which we modify}{, previous work including SparTen does not address the above scale-related issues  of feature map barrier avoidance, }{\em {intra}}{-filter load imbalance and on-chip bandwidth.}

Our simulations show that, on average, \name  performs  5.4x, 2.2x, 1.7x, and 2.5x better and achieves 19\%, 67\%, 7\%, and 7\% lower compute energy  than a dense, a one-sided, a naively-scaled two-sided architecture, and an iso-area two-sided architecture, respectively. Estimates from ASIC synthesis for four 8K-PE clusters using 45-nm technology estimates  1 GHz clock speed, 213 mm$^2$ area and 170 W power.
\putsec{background}{Background and challenges}

Each of CNN's many layers uses many filters to extract features~\cite{lecun,alexnet,vggnet,resnet}. The output feature map from one layer is the  input map to  the next layer. Each input map 
is a tensor with height $h$, width $w$, and depth $d$ which is 
the number of channels which, in turn, equals the number of filters in the  previous layer (\figref{cnn}).
Each filter is also a tensor  with typically equal height and width, $k$, and
the same depth as that of the  input map (\figref{cnn}). For a layer with $n$ filters, the output map dimensions are $h$ x $w$ x $n$. 
Each output map {\em cell}--  a scalar --  is  a  {\em tensor product} of the
input map and a filter (~\figref{cnn}). 
The filter ``slides over'' the input map  by a stride along the height and  width dimensions, one at a time,  in a two-dimensional convolution. 
The filter does not slide along the channel dimension. 
In a dense CNN, each layer performs   $h * w * k^2 * d * n$ multiply-adds
assuming  
a unit stride, ignoring boundary effects. Each filter (input map) is reused across all the input maps (filters).
Each filter cell in a channel is reused by every cell in the input map's corresponding channel ($h * w$ times). Similarly, each input map cell is reused for every filter ($k^2 * n$ times in all).

\figput{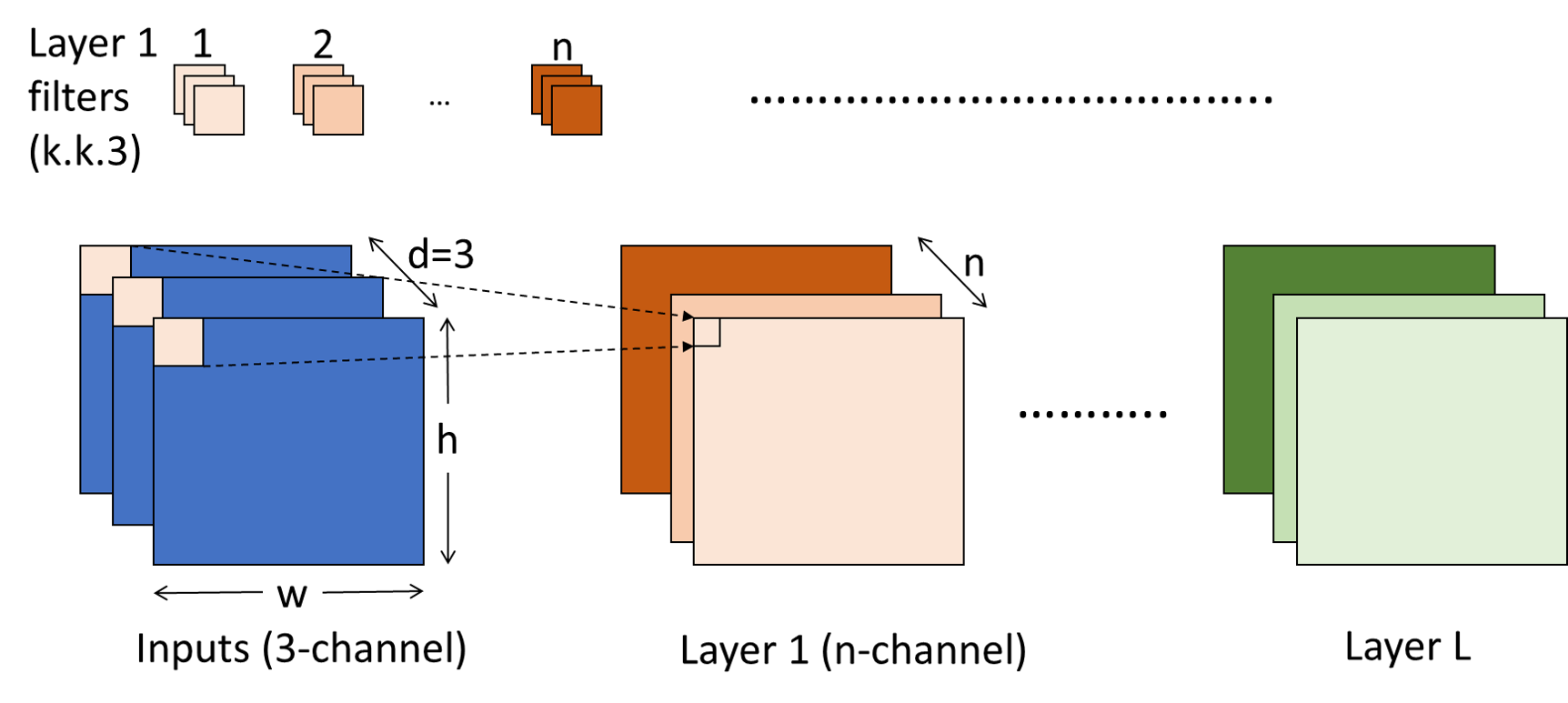}{}{Deep Convolutional  Neural Network}

Sparse CNNs  result in significantly less compute (e.g., 4-9x) and data (e.g., 2-3x) by avoiding zeros. In doing so, the reuse patterns remain the same though the reuse counts decrease. However,  implementing sparse vector-vector multiplication, the key primitive in sparse CNNs, is challenging.

\putsubsec{sparse-small}{Sparse CNN acceleration at small scale}

The central issue with small-scale two-sided sparse architectures is efficient implementation of the key primitive -- the sparse tensor-tensor multiplication. \figref{smallscale} shows a {\em processing element (PE)} which is the building 
block for  multiplication. For sparse tensors, the primitive involves finding the 
matching non-zero positions in the two tensors.  The sparse data representation impacts the efficiency of this matching. EIE~\cite{EIE} and SCNN~\cite{scnn}  employ Compressed  Sparse Row (CSR) 
whereas SparTen~\cite{sparten} employs a bit-mask representation.
CSR representation uses two vectors: one for the non-zero offsets and another for the non-zero values,
whereas the bit-mask representation uses a fixed-size bit mask to indicate zero and non-zero positions, and a vector for the non-zero values.  
The latter enables efficient matching via simple ANDing, prefix sum and priority encoder circuits.   ExTensor~\cite{extensor} proposes using CAMs for the matching.
SCNN avoids this matching altogether for unit-stride convolutions via an unusual dataflow  but incurs overheads resulting in intra- and inter-PE idling~\cite{sparten,bit-laconic}. 
In  SparTen, an output unit examines consecutive output channels, 
discards any zeros due to the Rectifier Linear Unit (ReLU) and writes the sparse output tensor, as per the chosen representation, to the on-chip cache or off-chip memory. 

In general, the architectures group many {\em lanes}, each of which includes a PE, into a {\em cluster} (\figref{smallscale}). 
In each cluster, each {\em lane} holds a different filter (lanes are analogous to our IFGCs). Each input map is broadcast
to the lanes to capture input map reuse across the filters. Multiple input maps are processed to capture filter reuse across the input maps. (Filter and input map reuse are symmetric and interchangeable.) The clusters may 
operate asynchronously with respect to each other. Alternatively, the filter or input map broadcast may cover multiple clusters. 

Buffering, load balancing, and on-chip bandwidth demand are key inter-related aspects of the architectures. The broadcasts save  on-chip bandwidth by avoiding repeated refetches of filters and input maps from the cache but require some buffering. While dense architectures like the TPU  can be efficient in buffering, sparsity makes such efficiency challenging. Dense CNNs are highly regular where every cell in a tensor is guaranteed to contribute  in a fixed pattern to the product. The TPU captures this regularity in a systolic architecture (proposed in~\cite{quantization}) where each MAC requires only a few buffers. Specifically, the TPU accumulates the partial product for a product along a column of the systolic array passing down the partial product to the next cell (near neighbor) achieving efficient pipelined parallelism within a tensor-tensor multiplication.  Thus, each MAC needs to buffer data only for a simple scalar-scalar multiplication, so that the MACs in a systolic column share the buffers needed for a product. 
Sparse CNNs, however, are irregular in that each non-zero scalar in a tensor contributes to the product only if there is a match in the other tensor. Consequently, the buffering in sparse accelerators has to increase to accommodate this uncertainty. 
Sparse accelerators cannot efficiently pipeline the multiplication due to the uncertainty, which disallows any buffer partitioning and forces the architectures to replace  intra-multiplication pipelined parallelism with inter-multiplication parallelism. While abundant, the latter disallows sharing of the buffers.

In addition to the buffering for the filters and input maps, the output maps also need buffering. However, because an entire tensor-tensor multiplication produces only a single cell in the output tensor, output buffers tend to be smaller than the filter and input map buffers.

While the filter and input map broadcasts save bandwidth, load imbalance across the lanes and clusters due to sparsity increases the need for  buffering the broadcast data. Due to uneven sparsity, a leading lane or cluster may run far ahead of the lagging  lane or cluster. 
Either all the broadcast data within this gap must  be  buffered or the broadcast must be stalled until the lagging entity has caught up enough to free up some buffers. 
Reasonable buffering often implies frequent stalls due to the implicit barriers imposed by the broadcasts in the presence of load imbalance.  A third option is 
to have the lagging entity refetch the broadcast data that could not be buffered. However, this option degrades  bandwidth.  Architectures exploiting bit sparsity ~\cite{bit-tactical} propose to alleviate load imbalance at the bit level by dynamically stealing work from adjacent lanes. However, such work-stealing involves dynamically scheduling and routing input and output values at high speeds which is complex to implement for full values. SparTen~\cite{sparten} proposes a load balancing scheme which either co-locates and serializes two filters on one PE  causing underutilization at large scales or employs a permutation network in each cluster that is hard to scale up (discussed in~\secref{inter-filter}).

\putsubsec{challenges}{Challenges in scaling up sparse CNN acceleration}

\figput[0.8\linewidth]{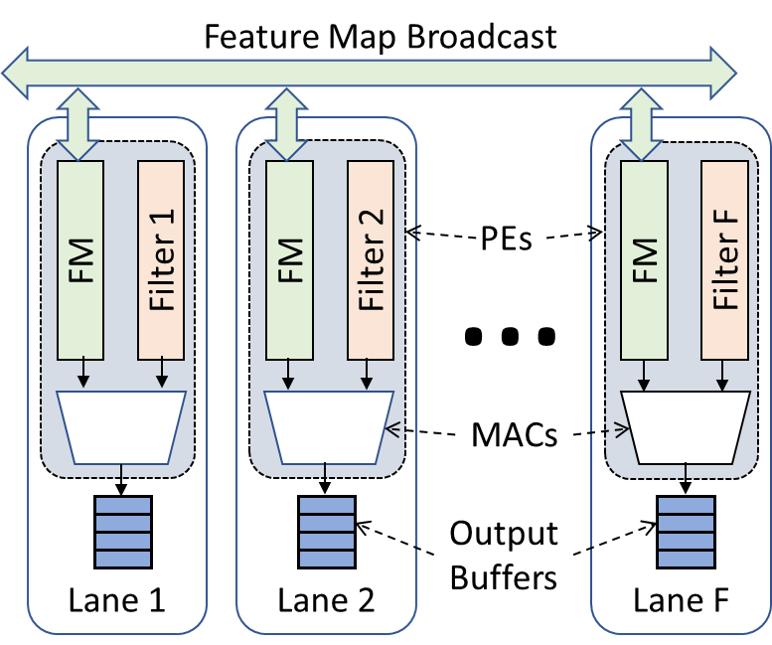}{}{A cluster in a small-scale sparse CNN accelerator}

At small scales (e.g., 1K MACs), broadcasting within a cluster and  employing asynchronous clusters  work  well with reasonable on-chip buffering (e.g., around 1-KB buffering per MAC amounts to only 1 MB total on-chip buffers). However,
the inter-related issues of buffering, load imbalance, and  bandwidth demand make it significantly challenging to scale up sparse architectures to the scales of dense architectures. For instance, naively scaling up SparTen's asynchronous  clusters with 32 MACs each to 32K MACs would require IK  clusters, requiring 1K refetches or more than 32-MB  on-chip buffering (in addition to a large cache).
On the other hand, synchronizing the clusters via broadcasts to limit the required bandwidth and buffering  would incur steep load imbalance at such extreme scales. Given that the filters are shared across the lanes (one axis) and the input  maps are shared across the clusters (another axis), {\em all} the MACs (e.g.,  32K at large scale) have to stay load-balanced to avoid excessive buffering or bandwidth demand.
\putsec{name}{BARISTA} 

Recall from~\secref{intro} that to address the above problems, \name proposes {\em barrier-free} execution  at large scale (i.e., no broadcasts). Note that barrier-free does not mean wait-free; \name incurs minimal waiting to achieve within 6\% of ideal performance. 
To prevent buffering and on-chip bandwidth requirements from exploding due  to  the  lack  of  broadcasts, 
\name
makes the following contributions: 
First,  to reduce the on-chip bandwidth 
\name employs  telescoping request combining for input maps and snarfing for filters (on-chip bandwidth demand). Second, to avoid barriers among the PEs between consecutive input maps per filter,  \name employs {\em output buffer coloring} so that processing of different input maps may overlap across the PEs (input feature map barrier avoidance).
Third,  to alleviate systematic load balance among a node's  PEs handling different filter parts,
\name employs simple {\em round-robin assignment of work} to the PEs (intra-filter load balance).
Further, to load balance across the filters, \name employs a variant of SparTen's Greedy Balancing (inter-filter load balance) in software.  A common theme among these techniques is that because of the scale they use either simple hardware or software.

Like previous architectures, \name exposes a simple interface for  matrix-vector  ($C \leftarrow Ax + y$) and matrix-matrix multiplications ($C \leftarrow A\times B)$. The interface linearizes tensors, which may be laid out non-contiguously in memory, into vectors for the relevant  operations.

\figput[\linewidth]{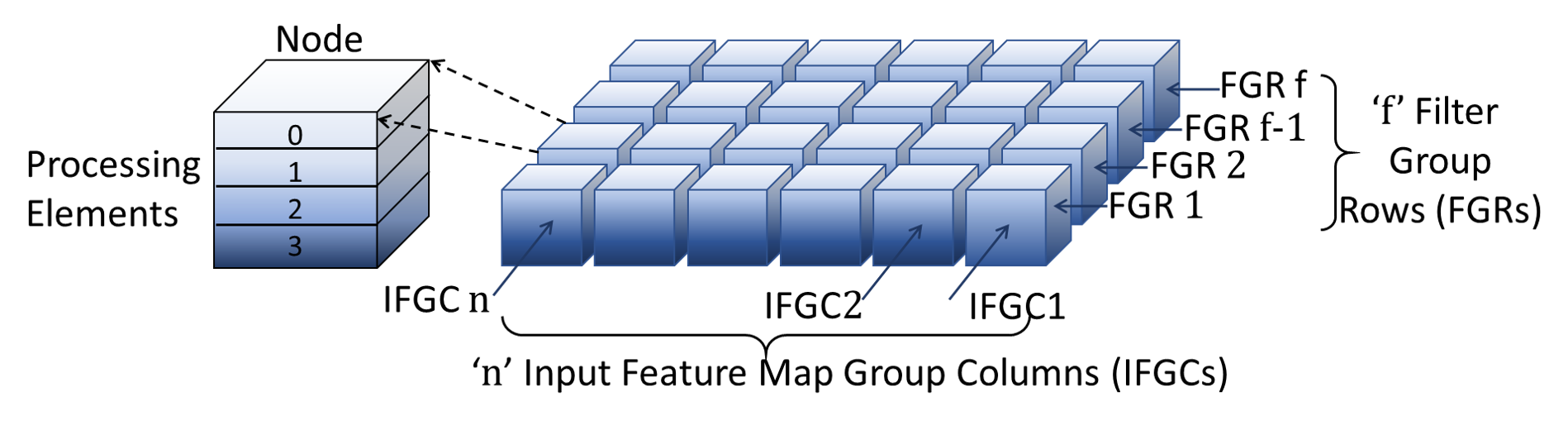}{}{Compute organization in \name}

\putsubsec{microarch}{BARISTA organization}  

To achieve good reuse of input maps and filters, \name employs much larger clusters. Each of  
\name's clusters   is organized as a three-dimensional compute grid of {\em filter group rows (FGRs)} and 
{\em input map group columns (IFGCs)} intersecting at nodes each of which comprises multiple {\em processing elements (PEs)} (\figref{tracksnlanes}). For instance, a cluster comprises 64 FGRs and 32 IFGCs  and  4 PEs per node  for a total of 8K PEs, achieving high within-cluster reuse. To match the TPU's 32K MACs (two clusters), \name requires four clusters which are connected to an on-chip cache.  Each FGR works mostly on one filter  at a  time (possibly straying to a few filters due to barrier freedom) and a different input map per node. Similarly, each IFGC works mostly on one input map at a time and a different filter per node.  
Employing multiple PEs per node reduces buffering by sharing a node's buffers among its PEs and improves load imbalance by requiring fewer IFGCs with lower spread of work variation. While the buffering decreases with more PEs per node, using  many  PEs would result in them running out of work due to sparsity. Assigning more work per IFGC to feed more PEs would leave some IFGCs idle (abundant parallelism but also vast compute resources).

Like other architectures, \name partitions filters and input maps into {\em chunks} for practical hardware granularity (e.g., 128 tensor cells).  To ensure that one output tensor cell is computed entirely at a node,
all the chunks of one full input map tensor and one full filter tensor are handled at the node. Processing of consecutive output channels  is assigned to consecutive nodes in an IFGC (\figref{tracksnlanes}). 
This assignment simplifies output handling, similar to previous proposals (e.g., ~\cite{eyeriss,sparten}).
Each input map chunk is processed by the filters in an IFGC (\figref{tracksnlanes}). Similarly, each filter chunk is processed by the input maps in an FGR. Thus, there is spatial reuse of each input map (filter) across the nodes in an IFGC (FGR). 
If there are more input maps  than the IFGCs, then the filters in an IFGC are reused over time across different input maps; similarly for more filters than FGRs.  The input map chunk and filter chunk  are  buffered at each node.
 
\figput{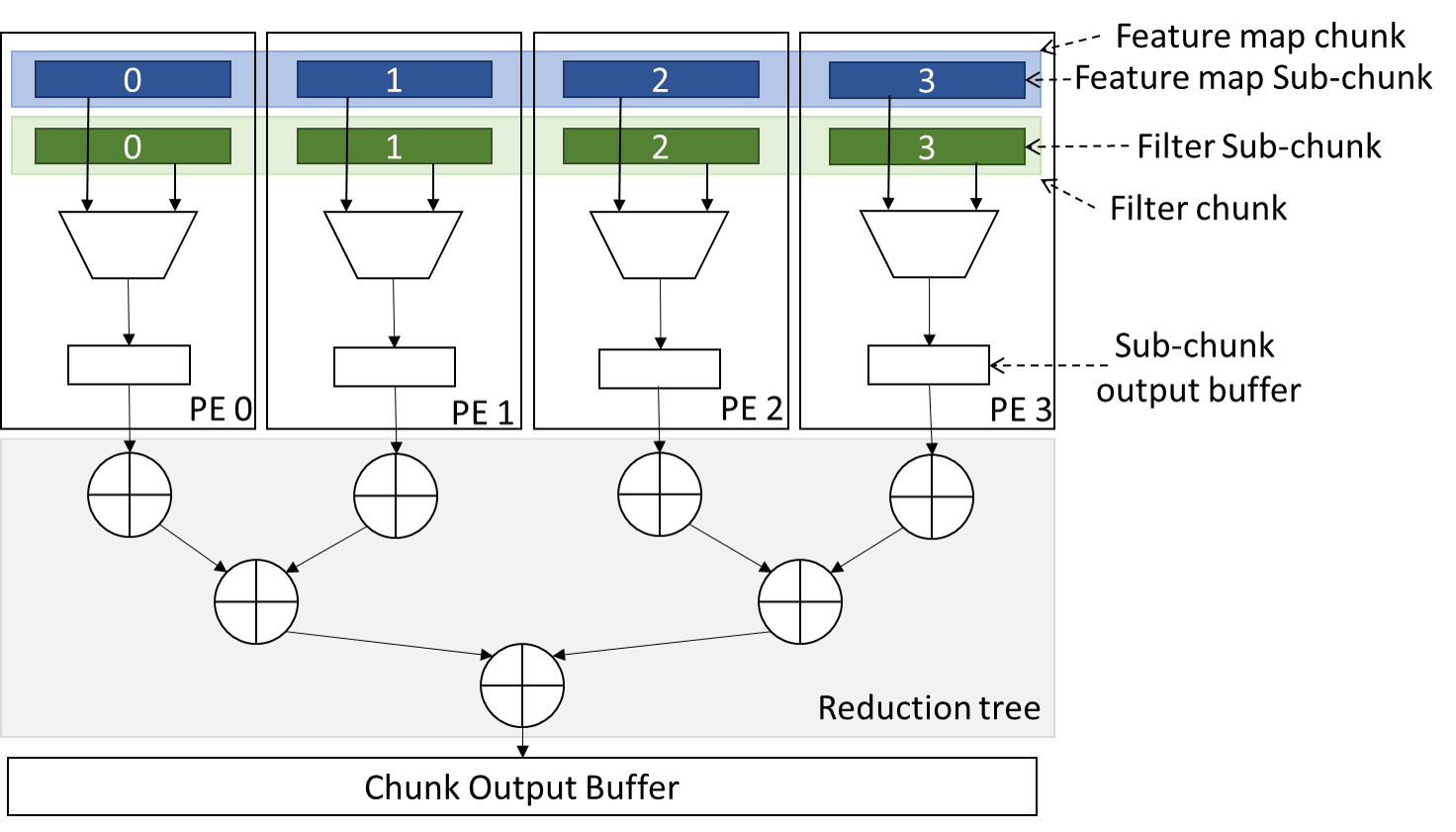}{}{Node organization}

At each node, the PEs {\em statically} partition the input map chunk into sub-chunks (\figref{node}). Each PE  multiplies its sub-chunk with the corresponding filter sub-chunk and accumulates the result in a {\em sub-chunk output buffer}. The sparse sub-chunk
multiplication may be accomplished by any of the previously-proposed methods (\secref{sparse-small}). Due to sparsity,
the PEs may complete at different times. When all of a node's  PEs are done, the  results are reduced through a small adder tree in the  node to the final {\em chunk output} which is buffered at the node (\figref{node}). When all the chunks of the tensors being multiplied are complete, the output cell is complete. 
When all the output cells in an IFGC are complete, a {\em conversion unit} forms the output in the chosen representation after discarding any zeros due to ReLU and writes back to an on-chip cache.
If there is temporal reuse of the same filter (input map) over several input maps (filters), then the chunk output for each input map (filter) is buffered separately at the node. 
A positive side-effect of the sub-chunks is that the PE circuitry (e.g., prefix sum and priority encode~\cite{sparten} or CAMs~\cite{extensor}) would be smaller than that for the full chunks.

The basic organization double-buffers the filters and input maps to hide latency.  However, if some nodes within IFGCs or FGRs are lagging and have not consumed their buffers, then each of those nodes would refetch the filters or inputs when its buffers are free. 
Next, we discuss our techniques for on-chip bandwidth demand,  load imbalance, and buffering.

\putsubsec{telescope}{Cutting on-chip bandwidth demand}

To eliminate the barriers induced by synchronous broadcasts, \name proposes to  employ asynchronous fetches of the input maps and filters. To reduce the ensuing bandwidth demand, \name employs  {\em snarfing} and {\em telescoping request combining}  for the filters and input maps, respectively. These schemes are based on the observation that if  the filters and input feature maps are reasonably load-balanced,  most of the nodes in an IFGC  would be working on the same input map chunk (with different filters). Similarly, most of the nodes in an FGR would be working  on the same filter chunk (with different input maps).  This in-sync progress implies that {\em even without (implicit) barriers}  the nodes in  an IFGC  would request the same input map chunks at about the same time; and the nodes in an FGR would request the same filter chunks at about the same time.  

Filters, being read-only for inference, can be load-balanced offline whose cost can be amortized over numerous inferences, but not input maps which are produced dynamically.
Assuming our variant of SparTen's  offline inter-filter load balancing (discussed later), the filters have much lower spread of density variation than the input maps.   Consequently, 
while the nodes in an IFGC stray less from each other in terms of progress,  there is more straying among the nodes in an FGR. Fortunately, a  common dichotomy in most architectures between filters and input maps is that each filter is reused  with many inputs both spatially (in each FGR)  and temporally (e.g., 16 times in each FGR node)  whereas each input is reused only spatially with many filters (in each IFGC). 
The roles of inputs and filters may be reversed but the dichotomy exists nevertheless because we hold one of either the filters or input maps and cycle through the other.  As such, the filters are reused more and hence are fetched less often. Therefore, the straying in an FGR does not result in excessive filter refetching. 

Due to this dichotomy, simple snarfing suffices for filters
so that when a node in an FGR makes a filter request, the other nodes that have empty filter buffers snarf the response (\figref{supstaBus}). 
The remaining nodes refetch the filter with the possibility of snarfing amongst themselves. The snarfing incurs only around  two refetches per filter due to the high filter reuse.

Though to a lesser extent than the FGR nodes, the nodes in an IFGC do stray. 
\figref{telescope} plots on the X axis  the progress of  input maps in an IFGC from AlexNet Layer 3 and the completion times on the Y axis. The filters are sorted by completion time. The figure shows the filters with two input maps, consecutive in time (the first input starts at time 20,000 and the second at 23,500).  
For the lower input map, a few fastest nodes are quite ahead whereas a majority of the  nodes stray only gradually from each other  followed in time  by fewer slower nodes. The rest of the figure is explained next. 

\figput{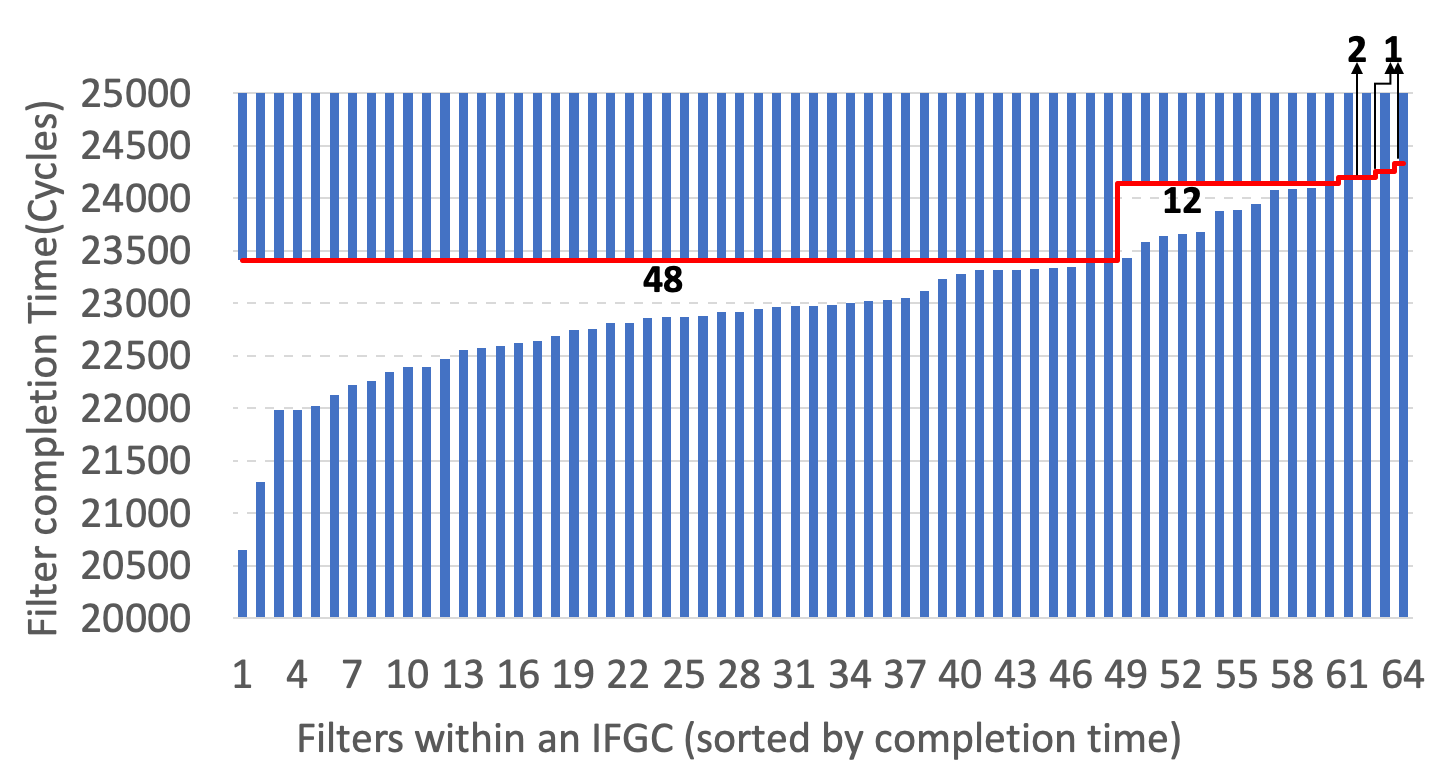}{}{Telescoping request combining}

For input map requests, simple snarfing would require many more refetches than those of the filters because of less reuse.
Instead, we observe that the IFGC nodes' requests that occur nearly together in time can be combined while introducing little delay. However, the straying implies that combining all the requests together would delay the leading nodes which may be the lagging nodes for later input maps. That is, with unlimited buffering, for instance, the leading nodes  would gain time with the current input maps and would spend the gains on future input maps for which the nodes lag, achieving good load balancing. Implicit barriers   between consecutive input maps, due either to synchronous broadcasts or to all-request combining,  prevent this transfer of time and hurt performance. Consequently, \name reaches a compromise by matching the request combining to the  tapering  nature of the number of nodes in the straying groups.  Instead of combining sets of the same number of requests, \name combines {\em telescoping} numbers of requests  (e.g., out of the 64 requests for the lower input map in~\figref{telescope}, \name  combines  the  first 48 requests, the next 12, the next two, and leaves the last two uncombined).  Even though this example implies five refetches,   often the requests in the next set arrive before the first set response increasing the effective combining count. Thus, in practice, the example configuration makes only three refetches on average. See~\figref{supstaBus} (the buffers are explained later). This scheme requires, per IFGC,  (a) a counter for the requests and  (b) a simple state machine to effect the telescoping count sequence. 

\putsubsec{load-balance}{Tackling load imbalance}  

While PEs, IFGCs, and FGRs all incur load imbalance, each IFGC operates on one input map  and multiple filters which may incur load imbalance.
As mentioned before, while filters are load-balanced offline (i.e., inter-filter load balance), input maps are not.  
Next, we address the {\em input map} load imbalance  and {\em intra-}filter load imbalance within a node.

\figput{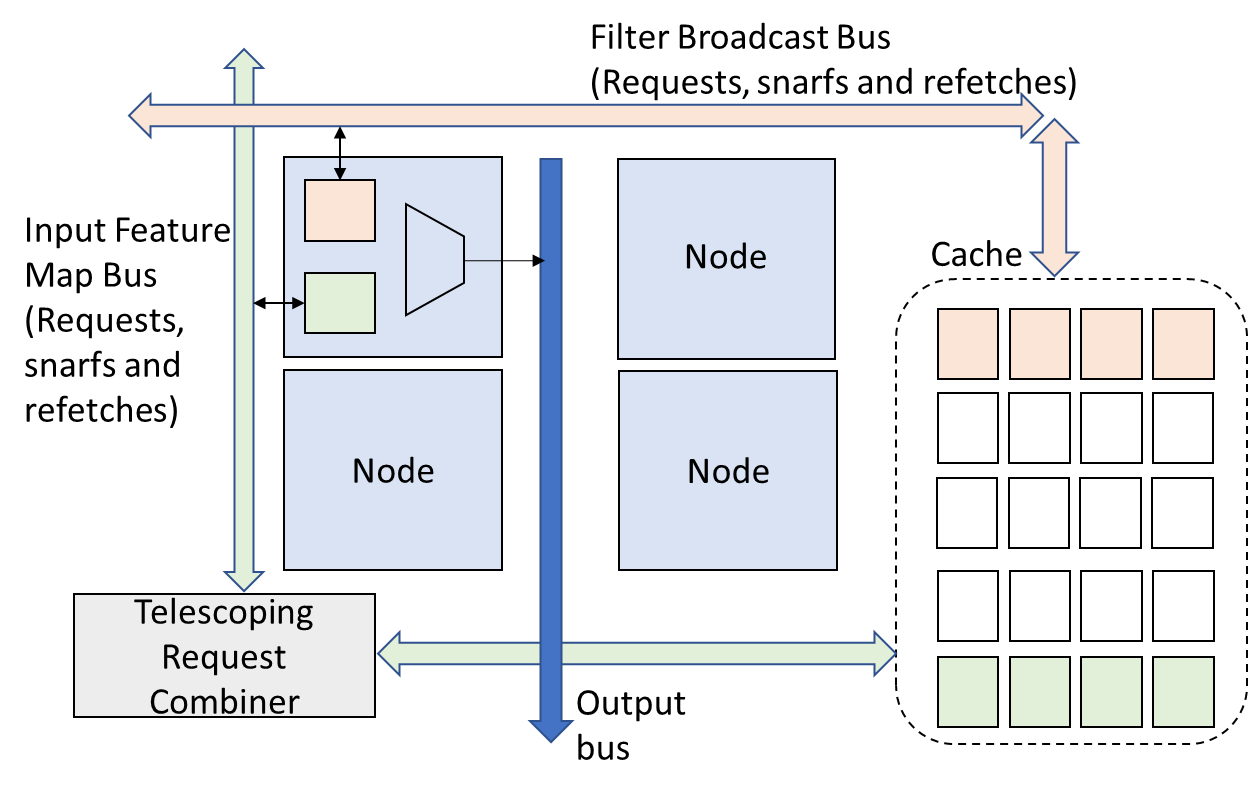}{}{Data transfer with snarfing and request combining}

\putsubsubsec{feature-map}{Input feature map barrier avoidance:}  

To achieve good filter reuse, a given  filter at a node in an FGR processes many input feature maps separated by node-local barriers. We avoid these
barriers which impose input  map load imbalance overhead. That is, some PEs in the node can move on to the next input map  while others are still working on the previous input map. Because the input maps are different, the chunk outputs are buffered separately. Previously, however, the input maps are processed sequentially without overlap so that all the PE's outputs go to the same output buffer at any given time. Now,  the PEs' outputs can 
concurrently go to different outputs. Accordingly, our coloring
scheme provides different sub-chunk output buffers for each PE (\figref{node}). Further, the coloring  matches the inputs with the sub-chunk output buffers using simple tags  (e.g., 4-bit tags for 16 input maps and output buffers).  Fortunately, the extra sub-chunk buffering cost is minimal because multiplying entire tensors produces just one output cell (e.g., 4 bytes per node). 

While coloring addresses input map load imbalance within a node, the imbalance across IFGCs is harder to address due to the dynamic nature of the input maps. Processing many input maps per filter for filter reuse (e.g., 16 input maps) helps to lessen this imbalance over a large amount of work.

\putsubsubsec{Intra-filter}{Intra-filter load balance:}  

While buffer sharing among a node's PEs reduces buffering, 
a major source of  load imbalance among the node's PEs is the static partitioning and assignment of a filter's sub-chunks to the PEs.
The   assignment  is susceptible to  systematic load imbalance across the PEs.  For instance, if a dense sub-chunk of a filter chunk is assigned to a PE to be reused for several input maps over time then the PE would lag another that is assigned a sparser sub-chunk. Here, we assume that a denser filter sub-chunk would result in more multiplication work than a sparser one (i.e., sub-chunk density is a proxy for multiplication work),
even though the work depends on the number of the matching non-zero positions in the filter and input map, and not the  number of  non-zero positions in the filter alone. 
The latter PE has to wait for the former before the input map buffer is released and the next input map  can be fetched (temporal reuse of the filter over several input maps). 
To address this issue, \name exploits the diversity in the density of sub-chunks of the input map chunks processed by an IFGC. Accordingly, \name employs  
{\em dynamic round-robin assignment of filter sub-chunks} to the PEs over consecutive input  maps chunks assuming  that the intra-chunk density distribution among the sub-chunks would even out over multiple chunks. For example. PE $i$ handles sub-chunk $i$ in chunk $0$, sub-chunk $i+1$ in chunk $1$, and so on. 
This assignment requires simple demultiplexing of the sub-chunks into the node's buffers (e.g., 1-4 demultiplexor for 4 PEs per node).

\putsubsubsec{inter-filter}{Inter-filter load balance:}  
The filters in an IFGC may vary in density causing load imbalance across the IFGC's nodes.
SparTen's {\em Greedy Balancing software variant (GB-S)}~\cite{sparten} addresses this load imbalance by offline sorting the whole  filters by density and co-locating  pairs of filters at each node, the densest and sparsest, the second densest and second sparsest, and so on. The sorting 
and co-location ensure that the  total work per node of the co-located filters  is similar across the IFGC's nodes.  Such sorting scrambles the output channels.  To match  the scrambled output channels with the  filters' weights in the next layer, this scheme  statically reorders the next layer's weights in software. Thus, the offline processing proceeds layer by layer,  occurs once, and  is amortized over all inferences. 
This software scheme works well at small scale but serializes the filter pairs at a node leading to idling of nodes at larger scales. SparTen also proposes a software-hardware hybrid variant ({\em GB-H}) which sorts the filters at the chunk granularity.
To reorder the output channels for each chunk, GB-H employs a per-cluster permutation network  which needs to support only low bandwidth for SparTen's small clusters but much higher bandwidths for \name's larger clusters. 

As such, \name employs a GB-S variant (hence, not claimed as a contribution) which performs whole-filter sorting but not co-location. 
As discussed in~\secref{buffering}, the increased  buffering  absorbs some of the non-systematic load  imbalance  due to the lack of chunk-level sorting.
However, the lack of co-location
causes systematic load imbalance. 
The nodes in an IFGC with the denser filters  lag behind those with the sparser filters. To address this issue, \name alternates the filter-to-node assignment in an IFGC between increasing and decreasing orders of filter density for consecutive input maps. 
This assignment produces two mutually-reverse orderings of the output channels for consecutive input maps (i.e., only two fixed permutations as opposed to any permutation in GB-H). To reorder the output channels, the conversion unit (\secref{microarch}) requires a 2-1 multiplexor to choose between the two orderings as opposed to GB-H's full permutation network.

\putsubsec{buffering}{Hierarchical buffering} 

Assuming the bit-mask representation (\secref{sparse-small}), the input map and filter each uses a 128-bit mask and a 128-byte block for data values.
If there were only one PE per node, the cluster would have 128 IFGCs and 64 FGRs for 8K PEs.
Assuming 128 IFGCs, the output data would be 128 1-byte cells. In a 128-PE FGR without coloring, the total buffering would be  ((128 bits + 128 bytes) * 2 (input and filter) + 1 byte (output))  * 128  (PEs) * 2 (double-buffering)  = 72.25 KB (i.e., 578 B per PE). For 64 FGRs in a cluster and four clusters, the total would be around 18 MB. Using four PEs to share a node's buffering  brings this total down to 4.5 MB.

This amount of buffering, however, exposes significant load imbalance especially because the PEs exploit intra-chunk parallelism  shrinking the compute time per chunk  (e.g., assuming an average 5x reduction in work due to sparsity, 128-bit mask and 4 PEs imply around 6 multiplications per chunk).  Such short compute times are susceptible to load imbalance due to even small variances among the IFGCs and FGRs.  Because the input maps are spatially-reused within the IFGCs  and the filters are spatially-reused within the FGRs, {\em all} the 32 IFGCs and 64 FGRs (i.e., 2K nodes in \name's cluster) need to stay load-balanced to avoid waiting due to unavailable buffers. This  requirement is challenging  due to the scale involved. 

In addition to load balancing (\secref{load-balance}), more buffering of the filters and input maps would help.
However, naively increasing  the buffering to recover performance results in buffering bloat (e.g., 8x). 
Because the filters are reused more than the input maps (the dichotomy in \secref{telescope}), the former need less buffering  (e.g., 3x instead of 8x). 
However, filter reuse requires more output buffering (e.g., 16x is typical).

To reduce the input map buffering, using narrower buffers degrades the cache bandwidth due to narrow fetches.
Instead, \name proposes {\em hierarchical buffering}, which enables both high cache bandwidth via chunk-wide fetches into a few chunk-wide, shared input map  buffers (at each IFGC) and low buffering by employing narrower sub-chunk-wide buffers at each node (see~\figref{supstaBus}). 
While the shared buffers' cost is amortized over the entire IFGC, the per-node buffers impose most of the overall buffering cost.
However, fewer per-node buffers cause more waiting in  telescoping used by  IFGC nodes  to cut input map refetches from the shared buffers to the per-node buffers. Fortunately, 
because the  shared buffers are shared only within an IFGC and not globally, the
buffers can afford some refetches from the nodes (e.g., 2-3 refetches per IFGC). Consequently, less than 8x per-node buffering for input maps  suffices (e.g., 3x instead of 8x). 
In a 64-node IFGC, the total buffering is [(128 bits + 128 bytes)  (input) * 16 (shared buffer depth)] + 
[(128 bits + 128 bytes) (filter) + (32 bits + 32 bytes) (input/PE) * 4 (PEs/node)] * 64 (nodes) * 3 (per-node buffering) + [1 byte (coloring sub-chunk output/PE)  * 4 (PEs/node) + 1 byte (chunk output/node)] * 64  (nodes) * 16 (output buffering)  = 61.25 KB (i.e., 245 B per PE). For 32 IFGCs per cluster and four clusters, the total is 7.66 MB.

\putsec{method}{Methodology}

To evaluate \name, we build a cycle-level simulator and synthesize an ASIC from a System Verilog implementation.

\tabput{benchmarks}
{\small
\centering
\begin{tabular}{|l|l|l|l|}
\hline
Bench- & \# layers  & filter & input map  \\
mark & & density & density \\
\hline
\hline
AlexNet & 5 & 0.368 & 0.473  \\ \hline
ResNet18 & 17 & 0.336 & 0.486 \\ \hline
Inception-v4 & 20* & 0.570 & 0.317 \\ \hline
VGGNet & 13 & 0.334 & 0.446 \\ \hline
ResNet50 & 49 & 0.421 & 0.384 \\ \hline

\end{tabular}

}{\normalsize Benchmarks (* 2 inception C modules)}

\noindent
{\bf Benchmarks:}
To obtain the sparse versions of the networks in~\tabref{benchmarks}, we apply
pruning and retraining to the networks' filters, as described by the original work~\cite{Dally-NIPS}. As recommended, we retrain the filters to try to recover  the original accuracy without pruning.  The resulting sparsity is in line with previously-reported results~\cite{Dally-NIPS}. We  use a mini-batch of 32, {\em int8} representation, 224x224 inputs from ImageNet~\cite{imagenet}, and  SparTen's bit mask-based sparse representation~\cite{sparten}.

\noindent
{\bf Simulated systems:} 
To isolate the basic  impact of two-sided sparsity,  we  simulate each of a dense accelerator, a one-sided sparse architecture (e.g., Cnvlutin), and three   two-sided sparse architectures --  SCNN (tile size 6x6), SparTen, and \name (see \tabref{hw-param}). We scale up the previous sparse architectures from their proposed scale by simply adding more clusters. To avoid underutilization in SCNN, each cluster operates on an independent image in the batch.
The main point of this paper is that achieving reuse via broadcasts induces implicit barriers which raise the inter-related issues of load imbalance, buffering and bandwidth demand.   To study this point we simulate other systems.  While SCNN employs  synchronous 
broadcasts across all clusters, SCNN incurs overheads related to its Cartesian product 
approach in addition to the barrier costs. 
To isolate the barrier cost of  broadcasts, we simulate  a broadcast-based scheme, 
called {\em Synchronous} which employs intra-cluster broadcasts (8K-MAC clusters similar to \name).  SparTen, which employs 
asynchronous refetches across its small clusters (and local broadcasts within each cluster), incurs high bandwidth at scale (numerous clusters). To isolate the impact of \name's optimizations, we simulate \name-no-opts which has \name's  organization but not the optimizations (refetches within its large clusters). 
To evaluate the amount of buffering needed to avoid barrier costs, we simulate a  broadcast scheme with unlimited buffering, called {\em Unlimited-buffer}.  Note that what matters is the granularity of the broadcasts; not the cluster size. Broadcasts to many PEs (within one large cluster or across many small clusters) expose  load imbalance or incur  buffering, whereas broadcasts to fewer PEs (in a subset of a large cluster or a small cluster) or no broadcasts at all incur on-chip bandwidth.

We simulate the one-sided architecture, SparTen and \name based on  SparTen's PEs and bit-mask representation using 128-byte chunks.
\name  uses 64 FGRs and 32 IFGCs to balance filter-reuse and feature-map reuse. Because this paper's goal is not a design-space exploration, we choose reasonable parameters for \name based on light exploration - the best parameters would only improve  \name.
\tabref{hw-param} lists the hardware parameters. Except for {\em unlimited-buffer}, all the architectures have similar resources, such as PEs, buffering, and off- and on-chip bandwidth, to isolate
the impact of the architectural differences without confounding resource disparities. As discussed in~\secref{sparse-small}, a TPU-like dense accelerator needs only 8 bytes of buffering  per MAC to achieve the highest performance. Adding more buffers will hurt the accelerator's energy without improving performance. Sparse architectures need more buffering than dense  architectures but smaller caches due to eliding zeros (around 2.5x smaller). 
Due to their regularity, dense architectures need only a few banks to achieve the highest performance. For energy, we use estimates based on our RTL implementation.

\tabput{hw-param}{
\small
\centering
\begin{tabular}{|l|l|l|l|}
\hline
Architecture & MACs/cluster & \# clusters & buffer/MAC \\ \hline
\hline
Dense  &  16K & 2  & 8 B   \\ \hline
One-sided  &  32 & 1K  & 819 B    \\ \hline
SCNN  & 1K & 32 & 1.63 KB	 \\ \hline
SparTen & 32 & 1K & 993 B \\ \hline
Synchronous & 8K & 4 & 993 B \\ \hline
\name & 8K & 4 & 245 B \\\hline
\name-no-opts & 8K & 4 & 245 B \\\hline
Unlimited buffer & 8K & 4 & infinite \\\hline
\hline
Architecture & cache size & \multicolumn{2}{|c|}{cache banks}\\ \hline
Dense & 24 MB & \multicolumn{2}{|c|}{8}\\ \hline
Sparse & 10 MB & \multicolumn{2}{|c|}{32}\\ \hline
\end{tabular}
}
{\normalsize Hardware parameters}

\noindent
{\bf Area, Power, Clock speed estimates from ASIC synthesis:} 
We synthesize an ASIC for our System Verilog implementation of one \name cluster using Synopsys's Design Compiler and the 45 nm technology FreePDK45 library~\cite{freepdk45}.  
Lacking a memory compiler,  the library synthesizes the buffers using inefficient flip-flops instead of SRAM arrays. To avoid this artificial overhead, we separately model the area and power of the buffers using CACTI 6.5~\cite{cacti6.5}. For CACTI power estimation, we conservatively assume the activity to be  one read and one write every cycle. 

\putsec{results}{Results}
Our results show the following: (1) performance comparison of \name against other schemes, (2) execution time breakdown for all the schemes to explain their performance, (3) isolation of the impact of \name's techniques, (4) energy comparison,  and  (5) area, power, and clock speed estimates from an ASIC synthesis of our RTL implementation.

\putsubsec{perf}{Performance}
We compare the performance of a dense architecture ({\em Dense}), a one-sided  sparse architecture (e.g., Cnvlutin) where only the input  maps are  sparse ({\em One-sided}),  SCNN (a two-sided sparse architecture), SparTen (a two-sided architecture), Synchronous (a two-sided architecture which employs broadcasts), \name, and Ideal which has unlimited bandwidth and buffering. 
In addition, because SparTen uses 1.9X more area than \name (\secref{asic-results}), we also include an iso-area variant of SparTen ({\em SparTen-Iso}). Both One-sided and SparTen asynchronously refetch the data without broadcasts.
Our key claim is that at the extreme scales of \name, achieving reuse via broadcasts  induces implicit barriers which raise the inter-related issues of load imbalance, buffering and bandwidth demand.  The barriers expose load imbalance (shown by Synchronous) whereas  (a)  asynchronous refetches would impose severe bandwidth costs (shown by SparTen and  later by \name-no-opts) and (b) buffering the broadcasts would require inordinate amount of buffering (evaluated next). Our comparison illustrates these issues.  

\figput{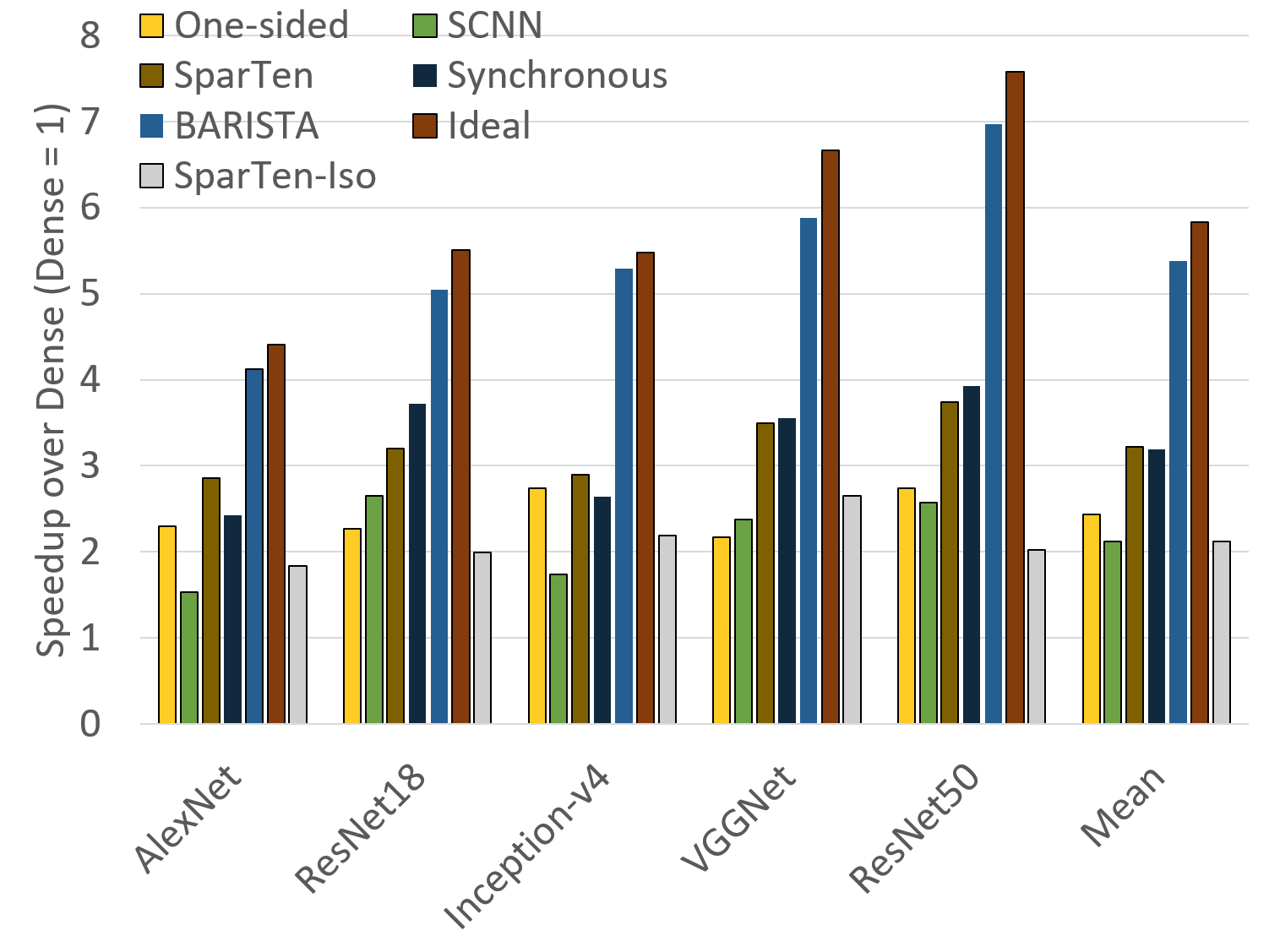}{}{Speedup}

\figref{speedup} shows the performance of these schemes normalized to that of {\em Dense} (Y-axis). The X-axis shows the benchmarks and their geometric  mean. To see the trends clearly, the benchmarks are ordered by increasing sparsity and hence opportunity (\tabref{benchmarks}).
One-sided improves over Dense by exploiting one-sided sparsity.  Though SCNN targets two-sided sparsity, its Cartesian product approach imposes significant overheads~\cite{bit-laconic,sparten}. These overheads drag SCNN behind One-sided in many benchmarks, despite SCNN targeting more sparsity than One-sided.  SparTen  improves over One-sided by exploiting two-sided sparsity while avoiding SCNN's overheads. SparTen's asynchronous clusters minimize barrier costs but impose significant bandwidth demand which Synchronous avoids via   broadcasts. However, Synchronous incurs implicit barriers  which push Synchronous slightly behind  SparTen on average, though the barrier costs are worse than the bandwidth penalty in some benchmarks and vice versa in others.

\name alleviates  both SparTen's bandwidth demand via snarfing and telescoping request combining (\name's bandwidth reduction schemes) and Synchronous's load imbalance via coloring, dynamic round robin, and the inter-filter load balancing variant (\name's load balancing techniques). Consequently, \name performs better than SparTen and Synchronous, achieving more than 5.4x average speedup over Dense and within 6\% of Ideal.  \name's speedup trends across the benchmarks match the sparsity (opportunity) trends. 

While \name performs 1.7x  better than SparTen, 
this comparison obscures the fact that SparTen, when scaled up to 32K MACs occupies significantly more area than \name (shown later in \secref{asic-results}). In an equal-area comparison  to SparTen-Iso, \name achieves 2.5x speedup.

\figput{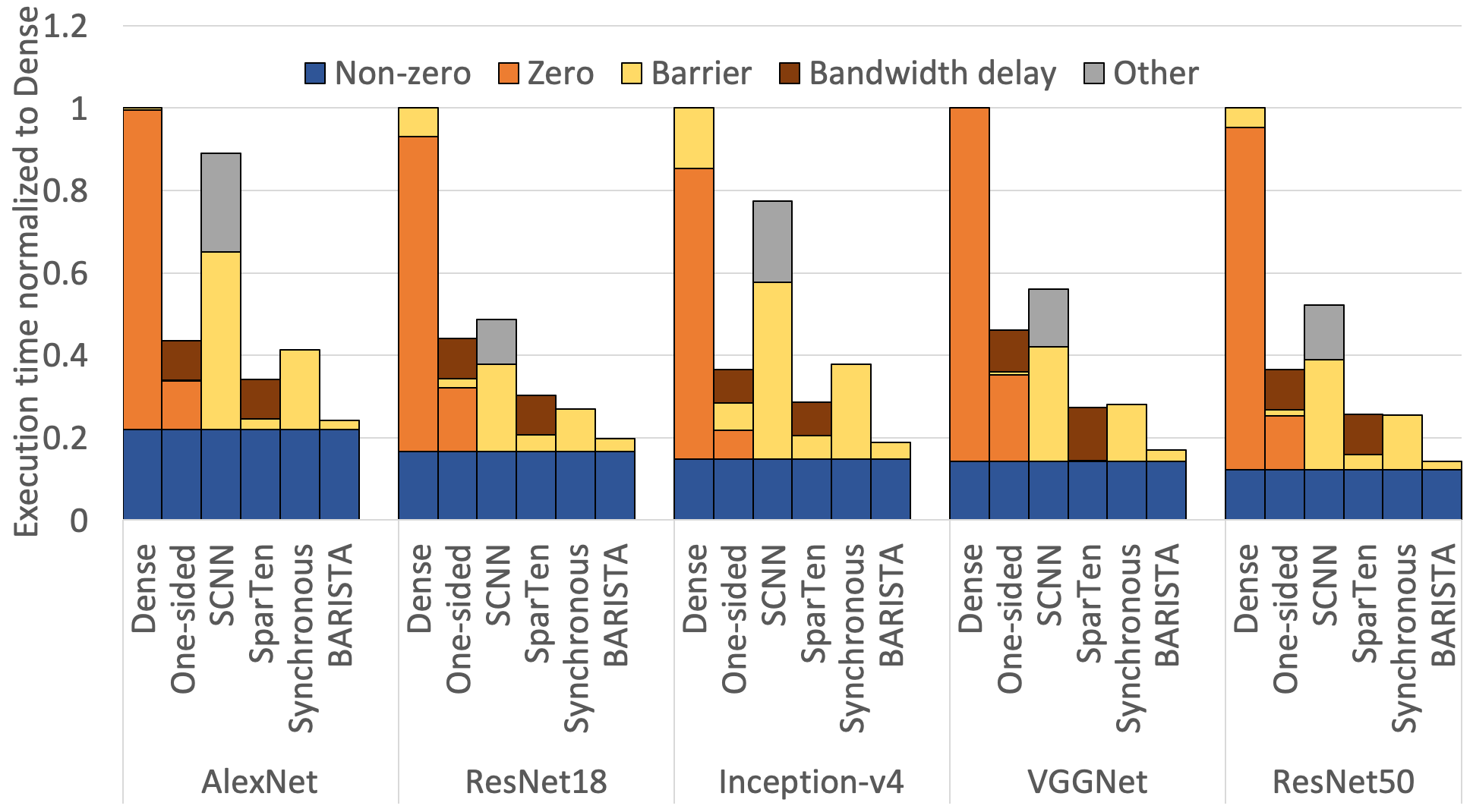}{}{Execution time breakdown}
 
To study the amount of buffering required to avoid both barrier costs and refetch bandwidth, we simulated \name by turning off the telescoping request combining and turning on unlimited buffering in a configuration called called {\em Unlimited-buffer} (\secref{method}). We found that in all the benchmarks Unlimited-buffer needs more than 24x buffering (i.e., more than 185 MB) to achieve the same performance  as \name.  

\putsubsec{breakdown}{Execution time breakdown}
To understand the speedups, \figref{breakdown} breaks down the above architectures' execution time for our benchmarks normalized to that of Dense. The execution time components are: non-zero computation, zero computation, barrier loss,  bandwidth-imposed delay, and other (explained below). As expected, Dense incurs many zero computations which is the motivation for sparse architectures. By exploiting one-sided sparsity One-sided incurs fewer zero computations than Dense but One-sided's refetches incur bandwidth delay.  SCNN exploits two-sided sparsity to eliminate zero computations but incurs significant overheads due to (1) its Cartesian product approach as mentioned above (shown as "other")  as well as (2) barriers induced by its broadcasts. 
The two-sided SparTen incurs low barrier cost due to local broadcasts only within its small clusters but high  bandwidth delays due to asynchronous refetches by the clusters. On the other hand, Synchronous incurs barrier cost due to the  broadcasts within its clusters. Finally, thanks to its bandwidth-reduction and load-balancing schemes, \name   incurs only some residual bandwidth delay and barrier cost. 

\putsubsec{energy}{Energy}
\figref{energy} shows the energy for Dense, Ones-sided, SparTen (with asynchronous refetches), and \name normalized to that of Dense on the Y axis, and the benchmarks and their mean on the X axis.  SCNN  performs worse than One-sided and is hard to model in enough detail for meaningful energy results. Therefore, we do not show SCNN. Further, our RTL synthesis tool chain does not estimate DRAM energy which cannot be normalized easily against the accelerator's energy. Therefore, we show compute and memory energies separately. We further break down (a) compute energy into zero, non-zero and data access (cache and buffers) components, and (b)  memory energy into zero and non-zero components. 

Dense's compute energy is dominated by zero compute and data access which includes both zeros and non-zeros. One-sided incurs higher compute energy than Dense because sparse computation requires more energy (finding the non-zero positions, as described in~\secref{sparse-small}). In the case of One-sided, this extra energy is incurred for both non-zeros and zeros (zeros are not elided in  one of the two input tensors). Though Dense incurs access energy for  both zeros and non-zeros, Dense achieves high reuse due to its regularity and therefore makes the fewest accesses possible (\secref{sparse-small}). On the other hand, One-sided both  accesses the zeros that are not elided, and  does not achieve as good reuse as Dense. As such, One-sided's clusters refetch data many times, incurring high data access energy (filters or input maps are symmetrical). Thus, One-sided's sparsity is not enough to reduce compute significantly but is enough to disrupt reuse and force refetches. 
 
\figputW[1.0\linewidth]{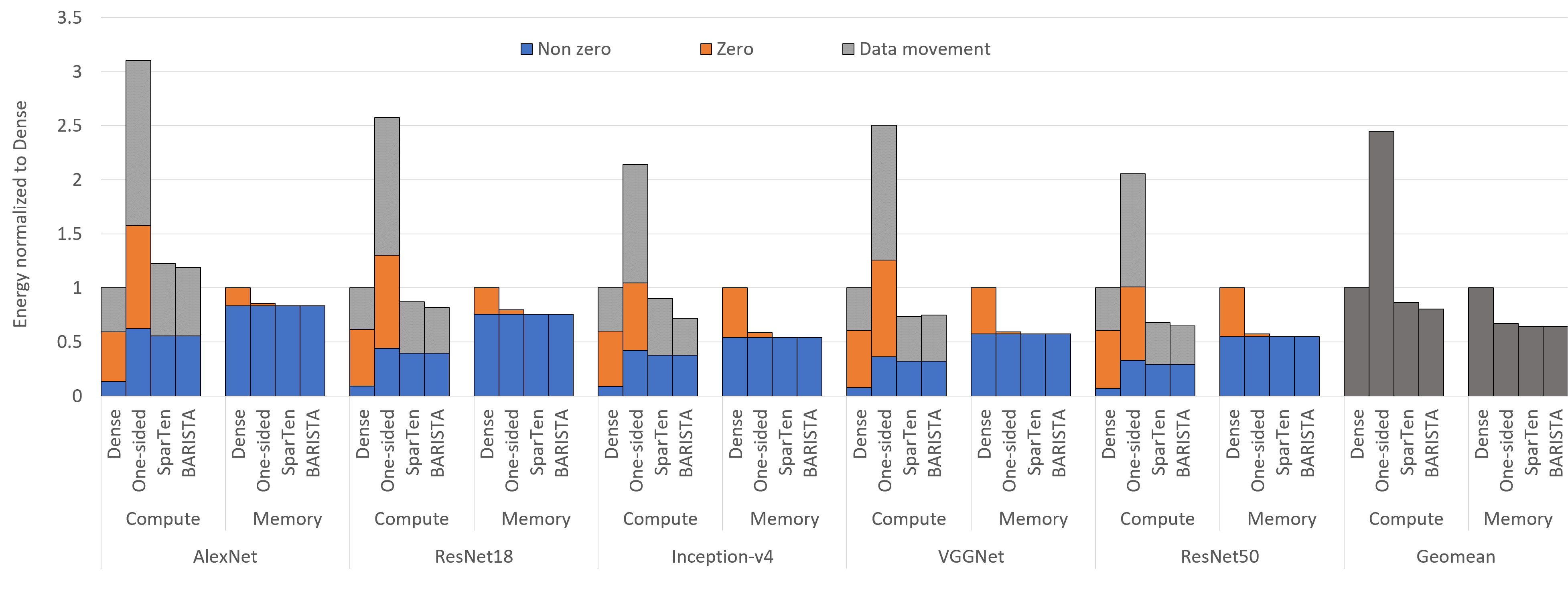}{}{Energy}

By exploiting two-sided sparsity, SparTen entirely avoids zero compute but incurs higher non-zero compute energy because two-sided sparse computation takes even more energy than one-sided sparse computation (finding the matching non-zero positions in the two input tensors as opposed to the non-zero positions in one  input tensor). Like One-sided, SparTen also incurs more data access energy than Dense. However, unlike One-sided, SparTen entirely avoids accessing zeros to achieve lower data access energy than One-sided.  As a result of the opposing effects of no zero compute or access but higher non-zero compute and access, SparTen's net compute  energy starts out being more than that of Dense at the left end  of the graph and ends up being less than that of Dense at the right end (i.e., as sparsity increases from left to right). Finally, \name's compute energy is identical to that of SparTen because they employ the same PE circuitry (\secref{microarch}). Further, \name achieves only slightly lower  data access energy than SparTen because the former's shared buffer energy offsets the latter's refetch energy. The  impact of SparTen's refetches in the form of bandwidth delay  causes a wider  gap  between  SparTen and \name in performance (\figref{speedup}) than in energy (\figref{energy}) because of the naturally synchronous (bursty) nature of the refetches.  The bursts cause significant queuing due to cache bank conflicts which \name avoids by controlling the refetches. Like SparTen,  \name's compute energy also decreases with increasing sparsity from left to right.

Unlike Dense's compute energy, its memory energy is dominated by non-zeros because  compute volume decreases quadratically over sparsity whereas memory volume decreases only linearly. 
One-sided reduces memory energy by eliminating input map zeros. In memory energy, input maps dominate filters which have high reuse and low memory traffic due to batching.  Consequently, the two-sided SparTen and \name improve only slightly over One-sided in memory energy despite eliminating both input map and filter zeros. Though One-sided, SparTen, and  \name incur some overhead in non-zero memory energy due to the bit masks and pointers for sparse data, the overhead is invisible. As the sparsity increases from left to right, the savings from zeros exceed the penalty from non-zeros so that  \name's  memory energy decreases.

\putsubsec{balancing}{Isolating \name's techniques}

\begin{figure}[t]
\begin{minipage}{\linewidth}
\footnotesize 
\begin{center}
\includegraphics[width=\linewidth]{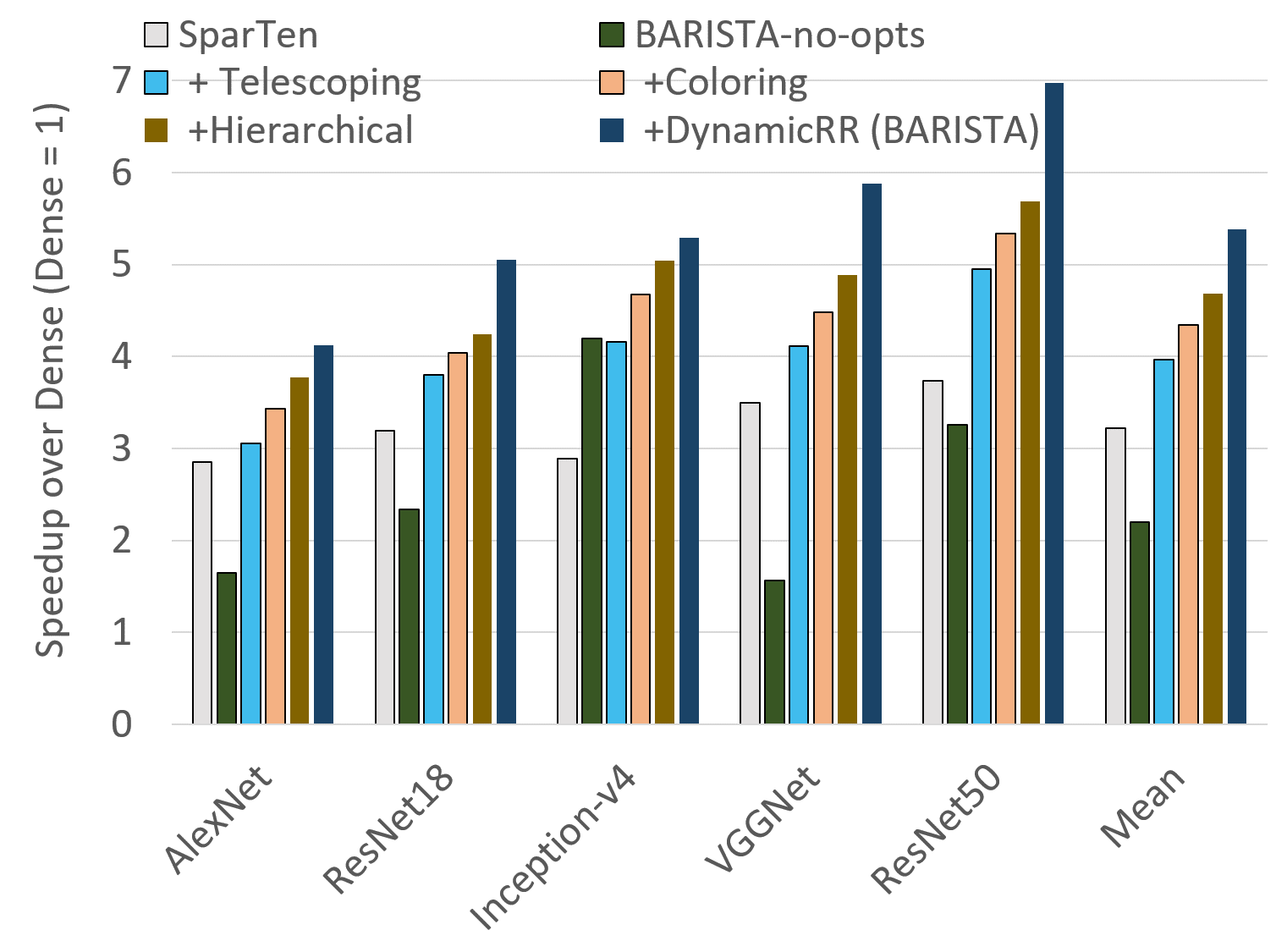}
\end{center}
\vspace{-0.25in}
\caption{Isolating \name's techniques \label{fig:isolation}}
\vspace{0.05in}
\end{minipage}
\end{figure}

\figref{isolation} isolates the impact of \name's techniques on performance. The Y axis shows the speedup over Dense for our benchmarks (including  their geometric mean).
The graph starts on the left  with  Sparten for reference and an initial \name configuration without any of the  optimizations (\name{\em-no-opts}). Like SparTen, this configuration (1) already includes SparTen's GB-S  for  inter-filter load balancing (the variant described in \secref{inter-filter})
and (2) employs asynchronous refetches which incur bandwidth delay.
To this 
configuration, we {\em progressively add one at a time} telescoping request combining,  coloring,  hierarchical buffering, and  dynamic round robin assignment  which results in the full \name design. Unsurprisingly,  the rise from Dense to SparTen  corresponds to exploiting sparsity, which is the central opportunity for all sparse architectures. 
Due to its higher on-chip bandwidth demand than SparTen, \name-no-opts performs worse than SparTen which while being asynchronous like \name-no-opts uses bandwidth-saving broadcasts {\em within} its clusters unlike \name-no-opts. 
Every one of \name's techniques contribute more or less similarly  to fill the gap between \name-no-opts and \name. The only exception is in {\em inception-v4} which has low data volume, and hence low in-chip bandwidth pressure, so that adding telescoping to \name-no-opts does not show any improvements.

\putsubsec{refetches}{Sensitivity to Buffer Size}

\name {'s snarfing and telescoping request combining are key optimizations for reducing the refetches due to the barrier-free 
straying of FGRs and IFGCs.} \figref{refetches} { shows refetches (Y-axis) averaged over all layers, and over all feature maps and filters for each benchmark (groups of bars along the X-axis) as we apply }\name {'s optimizations and as we vary the buffer size (bars within each group).
Without the optimizations (leftmost bar), the average number of refetches is dominated by the feature-map refetches; filter refetches, though included, are much fewer because filters are reused 16 times over multiple inputs. There is a dramatic drop in the number of refetches 
when }\name {'s optimizations are applied. Increasing buffer sizes (4 MB, 6 MB and 8 MB) result in progressively fewer refetches. We found that though bandwidth demand continues to drop with more buffering, there was no significant performance benefit beyond 8 MB (default).}

\figput{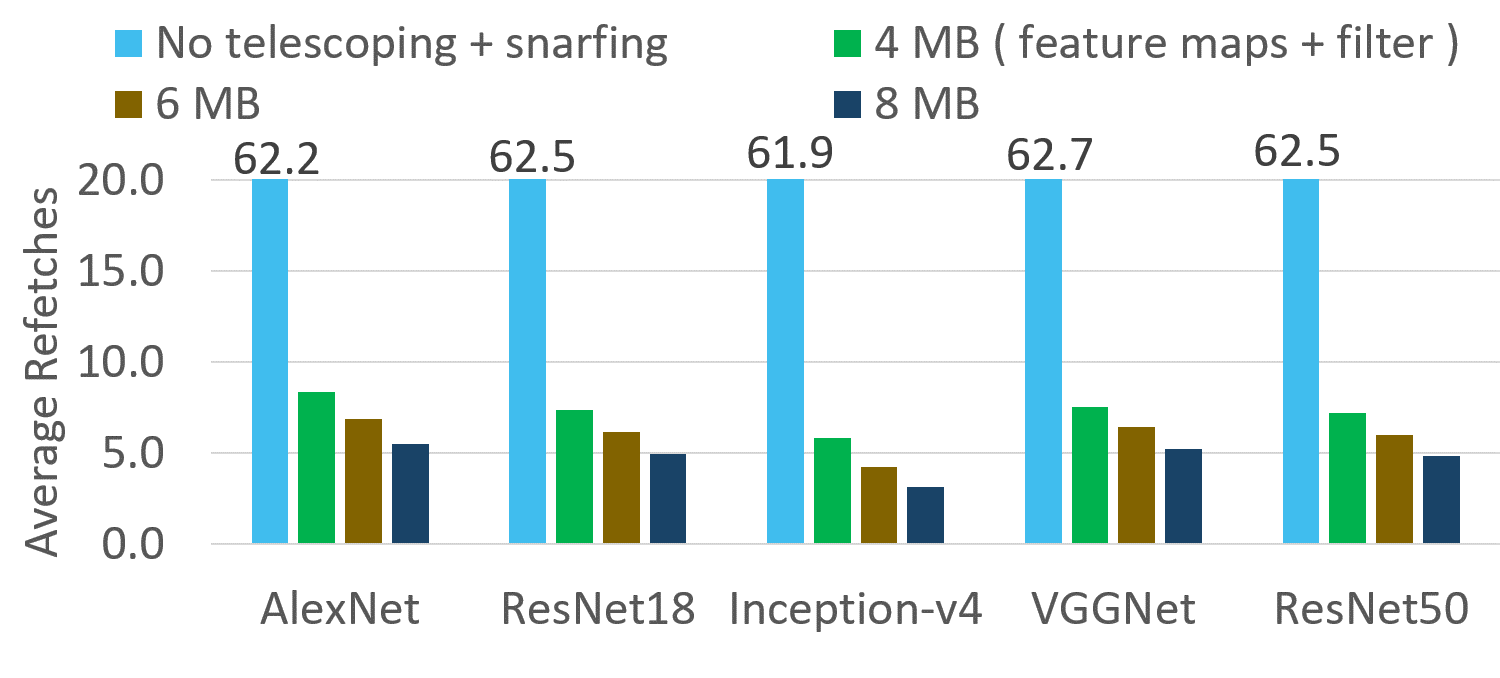}{}{Sensitivity to Buffer Size}

\putsubsec{asic-results}{ASIC synthesis results}

\tabref{areapower} shows the area and power estimates using 45-nm technology for four \name clusters with 8K PEs each. We scaled up  SparTen to 1K 32-PE clusters. 
These estimates show that \name's area and power are 89\% and 26\% smaller than those of SparTen; both designs achieve 1-GHz clock speed.
SparTen incurs higher  area and power due to larger buffers (\tabref{hw-param})
as well as control and bus interfaces replicated for 1K clusters. Otherwise, SparTen and \name have identical area and power for prefix sum, priority encoders and MACs, as expected. Prefix sum and priority encoders are smaller here for both \name and SparTen than in the original SparTen paper~\cite{sparten} due to sub-chunk-based circuits (32 B versus 128 B), as discussed in~\secref{microarch}.
In contrast, Dense achieves lower total area and power than \name due to smaller buffers and the lack of the sparsity circuitry. Dense's cache is larger to accommodate the non-sparse data.  
Overall, \name achieves 5.4x higher performance than Dense at the cost of 38\%  more area and 2.05x more power. 

\putsec{related}{Related work}

\tabput{areapower}{
\small
\begin{stripetabular}{ccccccc}
\footnotesize
\multirow[t]{2}{*}{} & \multicolumn{2}{c}{\name} & \multicolumn{2}{c}{SparTen} & \multicolumn{2}{c}{Dense} \\
 & Area  & Pwr  & Area  & Pwr & Area  & Pwr \\
 & (mm$^2$) &  (W) &  (mm$^2$) &  (W) & (mm$^2$) &  (W) \\
\hline
Buffers & 73.3 & 73.4 & 137.7 & 98.3 & 38.6 & 46.7 \\ 
Prefix & 43.6 & 43.1 & 43.6 & 43.1 & - & -\\
Priority  & 8.7 & 3.7 & 8.7 & 3.7 & - & - \\
MACs & 44.2 & 33.7 & 44.2 & 33.7  & 44.2 & 33.7 \\
Other  & 20.2 & 12.3 & 110.8 & 20.8 & 1.5 & 1.2 \\
\hline
Cache & 22.9 & 3.6 & 22.9 & 4.5 & 69.8 & 1.4 \\
\hline
Total & 212.9 & 170 & 402.7 & 214.9 & 154.1 & 83 \\
\end{stripetabular}
}
{\normalsize Area and power estimates for \name (45 nm)}

There are many optimizations for dense architectures, focusing on compute~\cite{snowflake, ferdman-fpga2, alexnet, tpu, brainwave, quantization}, memory~\cite{dadiannao, pudiannao, shidiannao}, and  reuse~\cite{eyeriss, fusion}. In-memory architectures~\cite{pim-Puma,pim-Pipelayer,pim-ISAAC,pim-Prime} leverage analog logic for  dense matrix multiplication but must contend with the well-known analog-related issues of noise, scalability, and process variation.
One-sided sparse architectures target sparsity  in either filters or feature maps but not both~\cite{cnvlutin,cambriconx,EIE,bit-tactical}
and two-sided architectures~\cite{scnn,sparten,extensor} exploit sparsity in the convolutional layers to reduce compute and data volumes. We have discussed SCNN and SparTen extensively. ExTensor~\cite{extensor} proposes hierarchical representations for sparse tensors.
EIE~\cite{EIE} targets two-sided sparsity only in the fully-connected layers but performs similarly to one-sided schemes due MAC idling for  zeros in the filters. Diffy~\cite{diffy} extends sparsity by exploiting small differences in the values. Other work proposes CPU instruction-set support   (a) to index efficiently into a hierarchical bitmap representation for sparse tensors~\cite{smash}
and (b) to store the data in memory after removing zeros while the computation does not elide zeros~\cite{zcomp}.

Bit-sparse architectures~\cite{bit-pragmatic,bit-tactical,bit-laconic}  leverage Booth encoding to elide zero bits. Unlike two-sided value-sparse architectures, the bit-sparse architectures incur overheads to store and transfer zero values.
Further, the fundamental issues of load imbalance, reuse, and  buffering remain for bit-sparse architectures as well (e.g., implicit barrier after every set of products in Bit Laconic and conservative buffering of uncompressed values before Booth encoding). 
PermDNN~\cite{permdnn} converts sparse filters into permuted diagonal matrices~\cite{banded}  only for fully-connected layers. 
CirCNN~\cite{circnn} uses block-circulant matrices for filters but requires
complex FFT hardware and does not capture all sparsity.

Other papers constrain sparsity to be coarse-grained to match hardware granularity~\cite{scalpel}, or by forcing zeros in contiguous~\cite{cambricons} or the same~\cite{Kung-CC} positions in groups of filters. In contrast, \name is based on Deep Compression~\cite{Dally-ICLR} which prunes each filter value independent of granularity and other filters. Such pruning carefully maintains accuracy by retraining whereas the constrained sparsity proposals either do not not evaluate accuracy on high-accuracy deep models~\cite{scalpel,cambricons} or lower accuracy (e.g., from 93.75\% to 93\%~\cite{Kung-CC} which is a 12\% increase in inaccuracy). 

In High-Performance Computing (HPC),  many implementations improve  sparse BLAS libraries' performance over CPUs~\cite{prasanna,HPCSparse}. However, the implementations remain significantly slower than dense accelerators like GPUs.

\putsec{concl}{Conclusion}

Scaling up two-sided sparse architectures 
is challenging  because alleviating the implicit barrier costs induced by broadcasts to achieve reuse raises 
the inter-related issues of load imbalance (e.g., keeping all 32K MACs load-balanced), buffering, and on-chip bandwidth demand. To address these issues, we proposed {\em barrier-free large-scale sparse tensor accelerator (\name)}.
\name (1) is the first architecture for scaling up sparse CNN accelerators; (2) reduces on-chip bandwidth demand by telescoping request combining the input map requests and   snarfing  the filter requests; (3) reduces buffering via basic buffer sharing and  avoids the ensuing barriers
between consecutive input maps by coloring the output buffers;  (4)  load balances a filter chunks among a node's PEs via dynamic round-robin work assignment; and  (5) employs hierarchical buffering for high cache bandwidth via wide fetches into shared buffers and narrower, private buffers at the compute. 
Our simulations show that, on average, \name  performs  5.4x, 2.2x, 1.7x, and 2.5x  better and achieves 
19\%, 67\%, 7\%, and 7\% lower compute energy 
than a dense, a one-sided, a naively-scaled two-sided, and an iso-area two-sided architecture, respectively.  Using 45-nm technology, ASIC synthesis  of our RTL implementation for four clusters of 8K PEs each reports 1-GHz clock speed, 213 mm$^2$ area and 170 W power.
We leave applying \name  to deep neural networks  other than CNNs and  sparse linear algebra in HPC  to future work. As such, \name is an attractive option for large-scale sparse tensor computation due to its at-scale high performance and efficiency.

\bibliographystyle{IEEEtranS}
\bibliography{local,cnn,gpu}

\begin{thebibliography}{10}
\providecommand{\url}[1]{#1}
\csname url@samestyle\endcsname
\providecommand{\newblock}{\relax}
\providecommand{\bibinfo}[2]{#2}
\providecommand{\BIBentrySTDinterwordspacing}{\spaceskip=0pt\relax}
\providecommand{\BIBentryALTinterwordstretchfactor}{4}
\providecommand{\BIBentryALTinterwordspacing}{\spaceskip=\fontdimen2\font plus
\BIBentryALTinterwordstretchfactor\fontdimen3\font minus
  \fontdimen4\font\relax}
\providecommand{\BIBforeignlanguage}[2]{{%
\expandafter\ifx\csname l@#1\endcsname\relax
\typeout{** WARNING: IEEEtranS.bst: No hyphenation pattern has been}%
\typeout{** loaded for the language `#1'. Using the pattern for}%
\typeout{** the default language instead.}%
\else
\language=\csname l@#1\endcsname
\fi
#2}}
\providecommand{\BIBdecl}{\relax}
\BIBdecl

\bibitem{agxorin}
``Nvidia introduces drive agx orin — advanced, software-defined platform for
  autonomous machines | nvidia newsroom,''
  \url{https://nvidianews.nvidia.com/news/nvidia-introduces-drive-agx-orin-advanced-software-defined-platform-for-autonomous-machines},
  (Accessed on 02/14/2020).

\bibitem{zcomp}
\BIBentryALTinterwordspacing
B.~Akin, Z.~A. Chishti, and A.~R. Alameldeen, ``Zcomp: Reducing dnn cross-layer
  memory footprint using vector extensions,'' in \emph{Proceedings of the 52Nd
  Annual IEEE/ACM International Symposium on Microarchitecture}, ser. MICRO
  '52.\hskip 1em plus 0.5em minus 0.4em\relax New York, NY, USA: ACM, 2019, pp.
  126--138. [Online]. Available:
  \url{http://doi.acm.org/10.1145/3352460.3358305}
\BIBentrySTDinterwordspacing

\bibitem{bit-pragmatic}
\BIBentryALTinterwordspacing
J.~Albericio, A.~Delmas, P.~Judd, S.~Sharify, G.~O'Leary, R.~Genov, and
  A.~Moshovos, ``Bit-pragmatic deep neural network computing,'' in
  \emph{Proceedings of the 50th Annual {IEEE/ACM} International Symposium on
  Microarchitecture, {MICRO} 2017, Cambridge, MA, USA, October 14-18, 2017},
  2017, pp. 382--394. [Online]. Available:
  \url{http://doi.acm.org/10.1145/3123939.3123982}
\BIBentrySTDinterwordspacing

\bibitem{cnvlutin}
\BIBentryALTinterwordspacing
J.~Albericio, P.~Judd, T.~H. Hetherington, T.~M. Aamodt, N.~D.~E. Jerger, and
  A.~Moshovos, ``Cnvlutin: Ineffectual-neuron-free deep neural network
  computing,'' in \emph{43rd {ACM/IEEE} Annual International Symposium on
  Computer Architecture, {ISCA} 2016, Seoul, South Korea, June 18-22, 2016},
  2016, pp. 1--13. [Online]. Available:
  \url{http://dx.doi.org/10.1109/ISCA.2016.11}
\BIBentrySTDinterwordspacing

\bibitem{fusion}
M.~Alwani, H.~Chen, M.~Ferdman, and P.~Milder, ``Fused-layer cnn
  accelerators,'' in \emph{49th Annual IEEE/ACM International Symposium on
  Microarchitecture (MICRO)}, 2016.

\bibitem{pim-Puma}
\BIBentryALTinterwordspacing
A.~Ankit, I.~E. Hajj, S.~R. Chalamalasetti, G.~Ndu, M.~Foltin, R.~S. Williams,
  P.~Faraboschi, W.-m.~W. Hwu, J.~P. Strachan, K.~Roy, and D.~S. Milojicic,
  ``Puma: A programmable ultra-efficient memristor-based accelerator for
  machine learning inference,'' in \emph{Proceedings of the Twenty-Fourth
  International Conference on Architectural Support for Programming Languages
  and Operating Systems}, ser. ASPLOS '19.\hskip 1em plus 0.5em minus
  0.4em\relax New York, NY, USA: ACM, 2019, pp. 715--731. [Online]. Available:
  \url{http://doi.acm.org/10.1145/3297858.3304049}
\BIBentrySTDinterwordspacing

\bibitem{HPCSparse}
N.~Bell and M.~Garland, ``Implementing sparse matrix-vector multiplication on
  throughput-oriented processors,'' in \emph{Proceedings of the Conference on
  High Performance Computing Networking, Storage and Analysis}, Nov 2009, pp.
  1--11.

\bibitem{eyeriss}
Y.-H. Chen, T.~Krishna, J.~Emer, and V.~Sze, ``14.5 eyeriss: An
  energy-efficient reconfigurable accelerator for deep convolutional neural
  networks,'' in \emph{2016 IEEE International Solid-State Circuits Conference
  (ISSCC)}, Jan 2016, pp. 262--263.

\bibitem{dadiannao}
\BIBentryALTinterwordspacing
Y.~Chen, T.~Luo, S.~Liu, S.~Zhang, L.~He, J.~Wang, L.~Li, T.~Chen, Z.~Xu,
  N.~Sun, and O.~Temam, ``Dadiannao: A machine-learning supercomputer,'' in
  \emph{Proceedings of the 47th Annual IEEE/ACM International Symposium on
  Microarchitecture}, ser. MICRO-47.\hskip 1em plus 0.5em minus 0.4em\relax
  Washington, DC, USA: IEEE Computer Society, 2014, pp. 609--622. [Online].
  Available: \url{http://dx.doi.org/10.1109/MICRO.2014.58}
\BIBentrySTDinterwordspacing

\bibitem{pim-Prime}
\BIBentryALTinterwordspacing
P.~Chi, S.~Li, C.~Xu, T.~Zhang, J.~Zhao, Y.~Liu, Y.~Wang, and Y.~Xie, ``Prime:
  A novel processing-in-memory architecture for neural network computation in
  reram-based main memory,'' in \emph{Proceedings of the 43rd International
  Symposium on Computer Architecture}, ser. ISCA '16.\hskip 1em plus 0.5em
  minus 0.4em\relax Piscataway, NJ, USA: IEEE Press, 2016, pp. 27--39.
  [Online]. Available: \url{https://doi.org/10.1109/ISCA.2016.13}
\BIBentrySTDinterwordspacing

\bibitem{TeslaFSD}
R.~Csongor, ``Tesla raises the bar for self-driving carmakers | the official
  nvidia blog,''
  \url{https://blogs.nvidia.com/blog/2019/04/23/tesla-self-driving/}, (Accessed
  on 02/14/2020).

\bibitem{bit-tactical}
\BIBentryALTinterwordspacing
A.~Delmas, P.~Judd, D.~M. Stuart, Z.~Poulos, M.~Mahmoud, S.~Sharify,
  M.~Nikolic, and A.~Moshovos, ``Bit-tactical: Exploiting ineffectual
  computations in convolutional neural networks: Which, why, and how,''
  \emph{CoRR}, vol. abs/1803.03688, 2018. [Online]. Available:
  \url{http://arxiv.org/abs/1803.03688}
\BIBentrySTDinterwordspacing

\bibitem{permdnn}
C.~{Deng}, S.~{Liao}, Y.~{Xie}, K.~K. {Parhi}, X.~{Qian}, and B.~{Yuan},
  ``Permdnn: Efficient compressed dnn architecture with permuted diagonal
  matrices,'' in \emph{2018 51st Annual IEEE/ACM International Symposium on
  Microarchitecture (MICRO)}, Oct 2018, pp. 189--202.

\bibitem{imagenet}
J.~Deng, W.~Dong, R.~Socher, L.-J. Li, K.~Li, and L.~Fei-Fei, ``{ImageNet: A
  Large-Scale Hierarchical Image Database},'' in \emph{CVPR09}, 2009.

\bibitem{circnn}
\BIBentryALTinterwordspacing
C.~Ding, S.~Liao, Y.~Wang, Z.~Li, N.~Liu, Y.~Zhuo, C.~Wang, X.~Qian, Y.~Bai,
  G.~Yuan, X.~Ma, Y.~Zhang, J.~Tang, Q.~Qiu, X.~Lin, and B.~Yuan, ``Circnn:
  Accelerating and compressing deep neural networks using block-circulant
  weight matrices,'' in \emph{Proceedings of the 50th Annual IEEE/ACM
  International Symposium on Microarchitecture}, ser. MICRO-50 '17.\hskip 1em
  plus 0.5em minus 0.4em\relax New York, NY, USA: ACM, 2017, pp. 395--408.
  [Online]. Available: \url{http://doi.acm.org/10.1145/3123939.3124552}
\BIBentrySTDinterwordspacing

\bibitem{shidiannao}
\BIBentryALTinterwordspacing
Z.~Du, R.~Fasthuber, T.~Chen, P.~Ienne, L.~Li, T.~Luo, X.~Feng, Y.~Chen, and
  O.~Temam, ``Shidiannao: Shifting vision processing closer to the sensor,'' in
  \emph{Proceedings of the 42Nd Annual International Symposium on Computer
  Architecture}, ser. ISCA '15.\hskip 1em plus 0.5em minus 0.4em\relax New
  York, NY, USA: ACM, 2015, pp. 92--104. [Online]. Available:
  \url{http://doi.acm.org/10.1145/2749469.2750389}
\BIBentrySTDinterwordspacing

\bibitem{brainwave}
\BIBentryALTinterwordspacing
J.~Fowers, K.~Ovtcharov, M.~Papamichael, T.~Massengill, M.~Liu, D.~Lo,
  S.~Alkalay, M.~Haselman, L.~Adams, M.~Ghandi, S.~Heil, P.~Patel, A.~Sapek,
  G.~Weisz, L.~Woods, S.~Lanka, S.~K. Reinhardt, A.~M. Caulfield, E.~S. Chung,
  and D.~Burger, ``A configurable cloud-scale dnn processor for real-time ai,''
  in \emph{Proceedings of the 45th Annual International Symposium on Computer
  Architecture}, ser. ISCA '18.\hskip 1em plus 0.5em minus 0.4em\relax
  Piscataway, NJ, USA: IEEE Press, 2018, pp. 1--14. [Online]. Available:
  \url{https://doi.org/10.1109/ISCA.2018.00012}
\BIBentrySTDinterwordspacing

\bibitem{banded}
\BIBentryALTinterwordspacing
N.~E. Gibbs, W.~G. Poole, Jr., and P.~K. Stockmeyer, ``A comparison of several
  bandwidth and profile reduction algorithms,'' \emph{ACM Trans. Math. Softw.},
  vol.~2, no.~4, pp. 322--330, Dec. 1976. [Online]. Available:
  \url{http://doi.acm.org/10.1145/355705.355707}
\BIBentrySTDinterwordspacing

\bibitem{snowflake}
V.~Gokhale, A.~Zaidy, A.~X.~M. Chang, and E.~Culurciello, ``Snowflake: An
  efficient hardware accelerator for convolutional neural networks,'' in
  \emph{2017 IEEE International Symposium on Circuits and Systems (ISCAS)}, May
  2017, pp. 1--4.

\bibitem{sparten}
\BIBentryALTinterwordspacing
A.~Gondimalla, N.~Chesnut, M.~Thottethodi, and T.~N. Vijaykumar, ``Sparten: A
  sparse tensor accelerator for convolutional neural networks,'' in
  \emph{Proceedings of the 52Nd Annual IEEE/ACM International Symposium on
  Microarchitecture}, ser. MICRO '52.\hskip 1em plus 0.5em minus 0.4em\relax
  New York, NY, USA: ACM, 2019, pp. 151--165. [Online]. Available:
  \url{http://doi.acm.org/10.1145/3352460.3358291}
\BIBentrySTDinterwordspacing

\bibitem{EIE}
S.~Han, X.~Liu, H.~Mao, J.~Pu, A.~Pedram, M.~A. Horowitz, and W.~J. Dally,
  ``Eie: Efficient inference engine on compressed deep neural network,'' in
  \emph{2016 ACM/IEEE 43rd Annual International Symposium on Computer
  Architecture (ISCA)}, June 2016, pp. 243--254.

\bibitem{Dally-ICLR}
\BIBentryALTinterwordspacing
S.~Han, H.~Mao, and W.~J. Dally, ``Deep compression: Compressing deep neural
  network with pruning, trained quantization and huffman coding,'' in \emph{4th
  International Conference on Learning Representations, {ICLR} 2016, San Juan,
  Puerto Rico, May 2-4, 2016, Conference Track Proceedings}, 2016. [Online].
  Available: \url{http://arxiv.org/abs/1510.00149}
\BIBentrySTDinterwordspacing

\bibitem{Dally-NIPS}
\BIBentryALTinterwordspacing
S.~Han, J.~Pool, J.~Tran, and W.~Dally, ``Learning both weights and connections
  for efficient neural network,'' in \emph{Advances in Neural Information
  Processing Systems 28}, C.~Cortes, N.~D. Lawrence, D.~D. Lee, M.~Sugiyama,
  and R.~Garnett, Eds.\hskip 1em plus 0.5em minus 0.4em\relax Curran
  Associates, Inc., 2015, pp. 1135--1143. [Online]. Available:
  \url{http://papers.nips.cc/paper/5784-learning-both-weights-and-connections-for-efficient-neural-network.pdf}
\BIBentrySTDinterwordspacing

\bibitem{resnet}
\BIBentryALTinterwordspacing
K.~He, X.~Zhang, S.~Ren, and J.~Sun, ``Deep residual learning for image
  recognition,'' \emph{CoRR}, vol. abs/1512.03385, 2015. [Online]. Available:
  \url{http://arxiv.org/abs/1512.03385}
\BIBentrySTDinterwordspacing

\bibitem{extensor}
\BIBentryALTinterwordspacing
K.~Hegde, H.~Asghari-Moghaddam, M.~Pellauer, N.~Crago, A.~Jaleel, E.~Solomonik,
  J.~Emer, and C.~W. Fletcher, ``Extensor: An accelerator for sparse tensor
  algebra,'' in \emph{Proceedings of the 52Nd Annual IEEE/ACM International
  Symposium on Microarchitecture}, ser. MICRO '52.\hskip 1em plus 0.5em minus
  0.4em\relax New York, NY, USA: ACM, 2019, pp. 319--333. [Online]. Available:
  \url{http://doi.acm.org/10.1145/3352460.3358275}
\BIBentrySTDinterwordspacing

\bibitem{tpu}
\BIBentryALTinterwordspacing
N.~P. Jouppi, C.~Young, N.~Patil, D.~Patterson, G.~Agrawal, R.~Bajwa, S.~Bates,
  S.~Bhatia, N.~Boden, A.~Borchers, R.~Boyle, P.-l. Cantin, C.~Chao, C.~Clark,
  J.~Coriell, M.~Daley, M.~Dau, J.~Dean, B.~Gelb, T.~V. Ghaemmaghami,
  R.~Gottipati, W.~Gulland, R.~Hagmann, C.~R. Ho, D.~Hogberg, J.~Hu, R.~Hundt,
  D.~Hurt, J.~Ibarz, A.~Jaffey, A.~Jaworski, A.~Kaplan, H.~Khaitan,
  D.~Killebrew, A.~Koch, N.~Kumar, S.~Lacy, J.~Laudon, J.~Law, D.~Le, C.~Leary,
  Z.~Liu, K.~Lucke, A.~Lundin, G.~MacKean, A.~Maggiore, M.~Mahony, K.~Miller,
  R.~Nagarajan, R.~Narayanaswami, R.~Ni, K.~Nix, T.~Norrie, M.~Omernick,
  N.~Penukonda, A.~Phelps, J.~Ross, M.~Ross, A.~Salek, E.~Samadiani, C.~Severn,
  G.~Sizikov, M.~Snelham, J.~Souter, D.~Steinberg, A.~Swing, M.~Tan,
  G.~Thorson, B.~Tian, H.~Toma, E.~Tuttle, V.~Vasudevan, R.~Walter, W.~Wang,
  E.~Wilcox, and D.~H. Yoon, ``In-datacenter performance analysis of a tensor
  processing unit,'' in \emph{Proceedings of the 44th Annual International
  Symposium on Computer Architecture}, ser. ISCA '17.\hskip 1em plus 0.5em
  minus 0.4em\relax New York, NY, USA: ACM, 2017, pp. 1--12. [Online].
  Available: \url{http://doi.acm.org/10.1145/3079856.3080246}
\BIBentrySTDinterwordspacing

\bibitem{smash}
\BIBentryALTinterwordspacing
K.~Kanellopoulos, N.~Vijaykumar, C.~Giannoula, R.~Azizi, S.~Koppula, N.~M.
  Ghiasi, T.~Shahroodi, J.~G. Luna, and O.~Mutlu, ``Smash: Co-designing
  software compression and hardware-accelerated indexing for efficient sparse
  matrix operations,'' in \emph{Proceedings of the 52Nd Annual IEEE/ACM
  International Symposium on Microarchitecture}, ser. MICRO '52.\hskip 1em plus
  0.5em minus 0.4em\relax New York, NY, USA: ACM, 2019, pp. 600--614. [Online].
  Available: \url{http://doi.acm.org/10.1145/3352460.3358286}
\BIBentrySTDinterwordspacing

\bibitem{alexnet}
\BIBentryALTinterwordspacing
A.~Krizhevsky, I.~Sutskever, and G.~E. Hinton, ``Imagenet classification with
  deep convolutional neural networks,'' in \emph{Advances in Neural Information
  Processing Systems 25}, F.~Pereira, C.~J.~C. Burges, L.~Bottou, and K.~Q.
  Weinberger, Eds.\hskip 1em plus 0.5em minus 0.4em\relax Curran Associates,
  Inc., 2012, pp. 1097--1105. [Online]. Available:
  \url{http://papers.nips.cc/paper/4824-imagenet-classification-with-deep-convolutional-neural-networks.pdf}
\BIBentrySTDinterwordspacing

\bibitem{Kung-CC}
\BIBentryALTinterwordspacing
H.~T. Kung, B.~McDanel, and S.~Q. Zhang, ``Packing sparse convolutional neural
  networks for efficient systolic array implementations: Column combining under
  joint optimization,'' \emph{CoRR}, vol. abs/1811.04770, 2018. [Online].
  Available: \url{http://arxiv.org/abs/1811.04770}
\BIBentrySTDinterwordspacing

\bibitem{cacti6.5}
H.~labs, ``Cacti 6.5,'' \url{https://www.hpl.hp.com/research/cacti/ }.

\bibitem{lecun}
Y.~Lecun, L.~Bottou, Y.~Bengio, and P.~Haffner, ``Gradient-based learning
  applied to document recognition,'' \emph{Proceedings of the IEEE}, vol.~86,
  no.~11, pp. 2278--2324, Nov 1998.

\bibitem{quantization}
\BIBentryALTinterwordspacing
D.~D. Lin, S.~S. Talathi, and V.~S. Annapureddy, ``Fixed point quantization of
  deep convolutional networks,'' in \emph{Proceedings of the 33rd International
  Conference on International Conference on Machine Learning - Volume 48}, ser.
  ICML'16.\hskip 1em plus 0.5em minus 0.4em\relax JMLR.org, 2016, pp.
  2849--2858. [Online]. Available:
  \url{http://dl.acm.org/citation.cfm?id=3045390.3045690}
\BIBentrySTDinterwordspacing

\bibitem{pudiannao}
\BIBentryALTinterwordspacing
D.~Liu, T.~Chen, S.~Liu, J.~Zhou, S.~Zhou, O.~Teman, X.~Feng, X.~Zhou, and
  Y.~Chen, ``Pudiannao: A polyvalent machine learning accelerator,'' in
  \emph{Proceedings of the Twentieth International Conference on Architectural
  Support for Programming Languages and Operating Systems}, ser. ASPLOS
  '15.\hskip 1em plus 0.5em minus 0.4em\relax New York, NY, USA: ACM, 2015, pp.
  369--381. [Online]. Available:
  \url{http://doi.acm.org/10.1145/2694344.2694358}
\BIBentrySTDinterwordspacing

\bibitem{diffy}
M.~{Mahmoud}, K.~{Siu}, and A.~{Moshovos}, ``Diffy: a d\'ej\`a vu-free
  differential deep neural network accelerator,'' in \emph{2018 51st Annual
  IEEE/ACM International Symposium on Microarchitecture (MICRO)}, Oct 2018, pp.
  134--147.

\bibitem{freepdk45}
NCSU, ``Freepdk45,''
  \url{https://www.eda.ncsu.edu/wiki/FreePDK45:Contents#Current_Version }.

\bibitem{scnn}
\BIBentryALTinterwordspacing
A.~Parashar, M.~Rhu, A.~Mukkara, A.~Puglielli, R.~Venkatesan, B.~Khailany,
  J.~Emer, S.~W. Keckler, and W.~J. Dally, ``Scnn: An accelerator for
  compressed-sparse convolutional neural networks,'' in \emph{Proceedings of
  the 44th Annual International Symposium on Computer Architecture}, ser. ISCA
  '17.\hskip 1em plus 0.5em minus 0.4em\relax New York, NY, USA: ACM, 2017, pp.
  27--40. [Online]. Available: \url{http://doi.acm.org/10.1145/3079856.3080254}
\BIBentrySTDinterwordspacing

\bibitem{vggnet}
\BIBentryALTinterwordspacing
O.~Russakovsky, J.~Deng, H.~Su, J.~Krause, S.~Satheesh, S.~Ma, Z.~Huang,
  A.~Karpathy, A.~Khosla, M.~S. Bernstein, A.~C. Berg, and F.~Li, ``Imagenet
  large scale visual recognition challenge,'' \emph{CoRR}, vol. abs/1409.0575,
  2014. [Online]. Available: \url{http://arxiv.org/abs/1409.0575}
\BIBentrySTDinterwordspacing

\bibitem{scaleout}
\BIBentryALTinterwordspacing
A.~Samajdar, Y.~Zhu, P.~N. Whatmough, M.~Mattina, and T.~Krishna, ``Scale-sim:
  Systolic {CNN} accelerator,'' \emph{CoRR}, vol. abs/1811.02883, 2018.
  [Online]. Available: \url{http://arxiv.org/abs/1811.02883}
\BIBentrySTDinterwordspacing

\bibitem{pim-ISAAC}
\BIBentryALTinterwordspacing
A.~Shafiee, A.~Nag, N.~Muralimanohar, R.~Balasubramonian, J.~P. Strachan,
  M.~Hu, R.~S. Williams, and V.~Srikumar, ``Isaac: A convolutional neural
  network accelerator with in-situ analog arithmetic in crossbars,'' in
  \emph{Proceedings of the 43rd International Symposium on Computer
  Architecture}, ser. ISCA '16.\hskip 1em plus 0.5em minus 0.4em\relax
  Piscataway, NJ, USA: IEEE Press, 2016, pp. 14--26. [Online]. Available:
  \url{https://doi.org/10.1109/ISCA.2016.12}
\BIBentrySTDinterwordspacing

\bibitem{bit-laconic}
\BIBentryALTinterwordspacing
S.~Sharify, A.~D. Lascorz, M.~Mahmoud, M.~Nikolic, K.~Siu, D.~M. Stuart,
  Z.~Poulos, and A.~Moshovos, ``Laconic deep learning inference acceleration,''
  in \emph{Proceedings of the 46th International Symposium on Computer
  Architecture}, ser. ISCA '19.\hskip 1em plus 0.5em minus 0.4em\relax New
  York, NY, USA: ACM, 2019, pp. 304--317. [Online]. Available:
  \url{http://doi.acm.org/10.1145/3307650.3322255}
\BIBentrySTDinterwordspacing

\bibitem{ferdman-fpga2}
Y.~Shen, M.~Ferdman, and P.~Milder, ``{Escher}: A {CNN} accelerator with
  flexible buffering to minimize off-chip transfer,'' in \emph{25th IEEE
  International Symposium on Field-Programmable Custom Computing Machines
  ({FCCM})}, 2017.

\bibitem{pim-Pipelayer}
L.~{Song}, X.~{Qian}, H.~{Li}, and Y.~{Chen}, ``Pipelayer: A pipelined
  reram-based accelerator for deep learning,'' in \emph{2017 IEEE International
  Symposium on High Performance Computer Architecture (HPCA)}, Feb 2017, pp.
  541--552.

\bibitem{scalpel}
\BIBentryALTinterwordspacing
J.~Yu, A.~Lukefahr, D.~Palframan, G.~Dasika, R.~Das, and S.~Mahlke, ``Scalpel:
  Customizing dnn pruning to the underlying hardware parallelism,'' in
  \emph{Proceedings of the 44th Annual International Symposium on Computer
  Architecture}, ser. ISCA '17.\hskip 1em plus 0.5em minus 0.4em\relax New
  York, NY, USA: ACM, 2017, pp. 548--560. [Online]. Available:
  \url{http://doi.acm.org/10.1145/3079856.3080215}
\BIBentrySTDinterwordspacing

\bibitem{cambriconx}
\BIBentryALTinterwordspacing
S.~Zhang, Z.~Du, L.~Zhang, H.~Lan, S.~Liu, L.~Li, Q.~Guo, T.~Chen, and Y.~Chen,
  ``Cambricon-x: An accelerator for sparse neural networks,'' in \emph{The 49th
  Annual IEEE/ACM International Symposium on Microarchitecture}, ser.
  MICRO-49.\hskip 1em plus 0.5em minus 0.4em\relax Piscataway, NJ, USA: IEEE
  Press, 2016, pp. 20:1--20:12. [Online]. Available:
  \url{http://dl.acm.org/citation.cfm?id=3195638.3195662}
\BIBentrySTDinterwordspacing

\bibitem{cambricons}
X.~{Zhou}, Z.~{Du}, Q.~{Guo}, S.~{Liu}, C.~{Liu}, C.~{Wang}, X.~{Zhou},
  L.~{Li}, T.~{Chen}, and Y.~{Chen}, ``Cambricon-s: Addressing irregularity in
  sparse neural networks through a cooperative software/hardware approach,'' in
  \emph{2018 51st Annual IEEE/ACM International Symposium on Microarchitecture
  (MICRO)}, Oct 2018, pp. 15--28.

\bibitem{prasanna}
\BIBentryALTinterwordspacing
L.~Zhuo and V.~K. Prasanna, ``Sparse matrix-vector multiplication on fpgas,''
  in \emph{Proceedings of the 2005 ACM/SIGDA 13th International Symposium on
  Field-programmable Gate Arrays}, ser. FPGA '05.\hskip 1em plus 0.5em minus
  0.4em\relax New York, NY, USA: ACM, 2005, pp. 63--74. [Online]. Available:
  \url{http://doi.acm.org/10.1145/1046192.1046202}
\BIBentrySTDinterwordspacing

\end{thebibliography}

\end{document}